\catcode`\@=11

\newif\if@fewtab\@fewtabtrue

{\count255=\time\divide\count255 by 60
\xdef\hourmin{\number\count255}
\multiply\count255 by-60\advance\count255 by\time
\xdef\hourmin{\hourmin:\ifnum\count255<10 0\fi\the\count255}}
\def\ps@draft{\let\@mkboth\@gobbletwo
    \def\@oddhead{}
    \def\@oddfoot
       {\hbox to 7 cm{$\scriptstyle Draft\ version:\ \draftdate$
       \hfil}\hskip -7cm\hfil\rm\thepage \hfil}
    \def\@evenhead{}\let\@evenfoot\@oddfoot}

\def\ceqno{\global\@fewtabfalse
    \ifcase\@eqcnt \def\@tempa{& & &}\or \def\@tempa{& &}
      \or \def\@tempa{&}
      \or\def\@tempa{}\fi\@tempa
{\rm(\theequation)}}

\def\aeqno#1{\global\@fewtabfalse
    \ifcase\@eqcnt \def\@tempa{& & &}\or \def\@tempa{& &}
      \or \def\@tempa{&}
      \or\def\@tempa{}\fi\@tempa
{\rm(\theequation,#1)}}

\def\label#1{\ifnum\draftcontrol=1
 \global\def\draftnote{$\scriptstyle #1$}\fi
 \@bsphack\if@filesw {\let\thepage\relax
   \def\protect{\noexpand\noexpand\noexpand}%
\xdef\@gtempa{\write\@auxout{\string
      \newlabel{#1}{{\@currentlabel}{\thepage}}}}}\@gtempa
   \if@nobreak \ifvmode\nobreak\fi\fi\fi
  \@esphack}

\def\alabel#1#2{\label{#1}\global\@fewtabfalse
    \ifcase\@eqcnt \def\@tempa{& & &}\or \def\@tempa{& &}
      \or \def\@tempa{&}
      \or\def\@tempa{}\fi\@tempa
{\hbox to 3cm{\phantom{\rm(\theequation,#2)}
\draftnote \hfil}\hskip -3cm {\rm(\theequation,#2)}}}

\def\clabel#1{\label{#1}\global\@fewtabfalse
    \ifcase\@eqcnt \def\@tempa{& & &}\or \def\@tempa{& &}
      \or \def\@tempa{&}
      \or\def\@tempa{}\fi\@tempa
{\hbox to 3cm{\phantom{\rm(\theequation)}
\draftnote \hfil}\hskip -3cm{\rm(\theequation)}}}

\def\eqnarray{\def\draftnote{{}}\global\@fewtabtrue
\stepcounter{equation}\let\@currentlabel=\theequation
\global\@eqnswtrue
\global\@eqcnt\z@\tabskip\@centering\let\\=\@eqncr
$$\halign to \displaywidth\bgroup\@eqnsel\hskip\@centering\@eqcnt\z@
  $\displaystyle\tabskip\z@{##}$&\global\@eqcnt\@ne
  \hskip 1\arraycolsep \hfil${##}$\hfil
  &\global\@eqcnt\tw@ \hskip 1\arraycolsep
$\displaystyle\tabskip\z@{##}$
\hfil  \tabskip\@centering&\global\@eqcnt\thr@@\llap{##}\tabskip\z@
\cr}

\def\endeqnarray{\@@eqncr\egroup
      \global\advance\c@equation\m@ne$$\global\@ignoretrue}

\def\@eqnnum{\hbox to 3cm{\phantom{\rm(\theequation)} \draftnote
                         \hfil}\hskip -3cm {\rm(\theequation)}}

\def\@@eqncr{\let\@tempa\relax
    \ifcase\@eqcnt \def\@tempa{& & &}\or \def\@tempa{& &}
      \or \def\@tempa{&}
      \or\def\@tempa{}
\fi\@tempa
\if@eqnsw
\if@fewtab\@eqnnum\fi
\stepcounter{equation}\fi\global
\@eqnswtrue\global\@eqcnt\z@\global\@fewtabtrue\cr}

\def\draftcite#1{\ifnum\draftcontrol=1#1\else{}\fi}

\def\@lbibitem[#1]#2{\item{}\hskip -3cm \hbox to 2cm
{\hfil$\scriptstyle\draftcite{#2}$}\hskip
1cm[\@biblabel{#1}]\if@filesw
     {\def\protect##1{\string ##1\space}\immediate
      \write\@auxout{\string\bibcite{#2}{#1}}}\fi\ignorespaces}

\def\@bibitem#1{\item\hskip -3cm \hbox to 2cm
{\hfil $\scriptstyle\draftcite{#1}$}\hskip 1cm
\if@filesw \immediate\write\@auxout
       {\string\bibcite{#1}{\the\value{\@listctr}}}\fi\ignorespaces}

\def\nsection#1{\section{#1}\setcounter{equation}{0}}

\def\nappendix#1{\vskip 1cm\no{\bf Appendix #1}\def\thesection{#1}
\setcounter{equation}{0}}

\font\tendl=msbm10  scaled \magstep1%double line
\font\sevendl=msbm7 scaled \magstep1
\font\fivedl=msbm5 scaled \magstep1
\font\tengl=eufm10  scaled \magstep1% gothic letters
\font\sevengl=eufm7 scaled \magstep1
\font\fivegl=eufm5 scaled \magstep1

\newfam\dlfam \def\dl{\fam\dlfam\tendl} % \dl is double line
\textfont\dlfam=\tendl \scriptfont\dlfam=\sevendl
\scriptscriptfont\dlfam=\fivedl
\newfam\glfam \def\gl{\fam\glfam\tengl} % \gl is gothic letters
\textfont\glfam=\tengl \scriptfont\glfam=\sevengl
\scriptscriptfont\glfam=\fivegl

\def\draftdate{\number\month/\number\day/\number\year\ \ \ \hourmin }

\global\def\draftcontrol{0}
\catcode`\@=12
\def\tilde{\widetilde}
\def\hat{\widehat}
\documentstyle[12pt]{article}

\def\theequation{{\thesection.\arabic{equation}}}

\newcommand{\be}{\begin{eqnarray}}
\newcommand{\en}{\end{eqnarray}\vs 0.5 cm}
\newcommand{\non}{\nonumber}
\newcommand{\no}{\noindent}
\newcommand{\vs}{\vskip}
\newcommand{\hs}{\hspace}

\newcommand{\p}{\partial}
\newcommand{\un}{\underline}
\newcommand{\NR}{{{\dl R}}}%letra doble raya en modo matematico
\newcommand{\NC}{{{\dl C}}}%letra doble raya en modo matematico
\newcommand{\NZ}{{{\dl Z}}}%letra doble raya en modo matematico
\newcommand{\NN}{{{\dl N}}}%letra doble raya en modo matematico
\newcommand{\qq}{\begin{eqnarray}}

\newcommand{\ee}{{\rm e}}

\newcommand{\qqq}{\end{eqnarray}}

\newcommand{\tr}{\hbox{tr}}

\newcommand{\CA}{{\cal A}}

\newcommand{\CC}{{\cal C}}
\newcommand{\CD}{{\cal D}}

\newcommand{\CG}{{\cal G}}

\newcommand{\CL}{{\cal L}}
\newcommand{\CM}{{\cal M}}

\newcommand{\CO}{{\cal O}}
\newcommand{\CP}{{\cal P}}
\newcommand{\CQ}{{\cal Q}}

\newcommand{\CS}{{\cal S}}
\newcommand{\CT}{{\cal T}}

\newcommand{\s}{\hspace{0.05cm}}
\begin{document}
\
\begin{center}
\vskip 0.9cm

%\Large{\bf{CHERN-SIMONS STATES AT GENUS ONE}}
\Large{\bf{Chern-Simons states at genus one}}
\vskip 1.2cm

\large{Fernando Falceto}\footnote{address after October 1, 1992: Depto.
F\'{\i}sica Te\'orica, U. Zaragoza, 50009 Zaragoza, Spain}

\large{Krzysztof Gaw\c{e}dzki}
\vskip 0.2cm
\large{C.N.R.S., I.H.E.S.,
91440 Bures-sur-Yvette, France}
\end{center}
\date{ }
%\maketitle

%%% for draft versions, suppress in definitive version
%\draft
%%
\vskip 1.2cm

\begin{abstract}
\noindent
%We consider the Schr\"{o}dinger picture quantization of
%the $SU(2)$ Chern-Simons theory in the toroidal geometry. A rigorous
%analysis of the spaces of states, of their parallel transport under
%the Knizhnik-Zamolodchikov-Bernard connection and of their
%dimension is presented.
We present a rigorous analysis of the Schr\"{o}dinger picture quantization
for the $SU(2)$ Chern-Simons theory on  3-manifold torus$\times$line, with
insertions of Wilson lines. The quantum states, defined as gauge covariant
holomorphic functionals of smooth
$su(2)$-connections on the torus, are expressed
by degree $2k$ theta-functions satisfying
additional conditions. The conditions
are obtained by splitting the space of semistable $su(2)$-connections into
nine submanifolds and by analyzing the behavior of states at four
codimension $1$ strata. We construct the Knizhnik-Zamolodchikov-Bernard
connection allowing to compare the states for different complex structures
of the torus and different positions of the Wilson lines. By letting
two Wilson lines come together, we prove a recursion relation for the
dimensions of the spaces of states which, together with the (unproven)
absence of states for spins$\s>{_1\over^2}$level implies the Verlinde
dimension formula.
\end{abstract}
\vskip 0.8cm

\nsection{\hspace{-.7cm}.\ \ Introduction}

Since the Chern-Simons (CS) theory was
revisited by E. Witten in \cite{Witten}, with the stress on its topological
nature, a considerable effort has been made to study different
aspects of the theory. From the point of view of covariant
quantization, the CS model was used to obtain 3-manifold, knot
and link invariants, either by surgery
\cite{Witten}\cite{Jeffrey} or in the perturbative expansion
\cite{GMM}\cite{GMM1}\cite{Bar-N}.

Here we will discuss a complementary aspect of the CS theory:
its canonical quantization. It has been argued in \cite{Witten}
that the space of Schr\"odinger states of
the CS theory with a compact Lie group $G$, in the presence of Wilson lines,
is isomorphic to the space of conformal blocks of the associated group $G$
Wess-Zumino-Witten (WZW) conformal field theory. The Wilson lines correspond
to the insertions of the primary fields in the two-dimensional
model. The conformal blocks are
holomorphic sections of the Friedan-Shenker (FS)
\cite{FS} vector bundle  over the moduli space of complex structures
of a punctured Riemann surface. By definition, they are horizontal with
respect to a flat projective connection. In the WZW model, the
latter is a generalization \cite{Denis1}\cite{Denis2}\cite{Witten}
\cite{Hitchin}\cite{Gawedzki}
\cite{AxelWitten}
of the genus zero Knizhnik-Zamolodchikov (KZ)
connection \cite{KnizhZamo}\cite{Kohno}.

For the case with no Wilson lines, the argument of \cite{Witten}
goes like follows. Working on a 3-manifold of the form $\Sigma\times\NR$,
with $\Sigma$ a 2-surface and $\NR$ interpreted as time, one identifies
the phase space of the Chern-Simons theory as the moduli space of flat
$G$-connections on $\Sigma$. The choice of a complex structure on $\Sigma$
allows to replace the latter with the moduli space of holomorphic
$G^{\NC}$-bundles on $\Sigma$ \cite{NarSesh}\cite{Donald}.
It determines, in the geometric quantization
jargon, a complex polarization of the phase space and, consequently,
a concrete realization of the space of CS states:
they become holomorphic sections of a power $\sim$ to the coupling
constant (level) $k$ of the determinant
line bundle over the moduli space of $G^{\NC}$-bundles. Such sections
form the fiber of the FS bundle for the WZW model with
the point in the base given by the complex structure of $\Sigma$.
The KZ connection in the FS bundle allows to compare
the CS states for different polarizations.

The presence of the Wilson lines parallel to the time axis, carrying
representations $R_n$ of $G$, leads to a slight modification
of the above picture. If we first treat the insertions classically
then the flat $G$-connections forming  the phase space should have
prescribed holonomies (up to conjugacy) around the punctures and, given
the complex structure of $\Sigma$, one obtains the moduli
space of holomorphic $G^{\NC}$-bundles on $\Sigma$ with parabolic
structure at the punctures \cite{MehSesh}. The CS states are
holomorphic sections of a power ($\sim k$) of the determinant line
bundle over the new moduli space.

Our approach will be somewhat different from the above.
First, one may treat the Wilson lines quantum-mechanically
from the start. The space of states which results then is composed of
holomorphic sections of a finite-dimensional vector bundle over
the old moduli space of $G^{\NC}$-bundles on $\Sigma$ (without parabolic
structures). The fiber of the bundle is the tensor product
of the representation spaces $V_{R_n}$, one for each Wilson line.

The two approaches should be equivalent provided the subtle
point of the behavior of the sections around singular points
of the moduli spaces is treated appropriately (stability, etc.).
We shall essentially follow the second approach
with another modification which allows precision and avoids at the
onset the subtleties inherent in the finite-dimensional
moduli-space context at a relatively low cost (one has to work with
infinite-dimensional Fr\'{e}chet spaces). Namely, we shall define
the CS states as holomorphic maps from the space of smooth
$G$-connections on $\Sigma$ to $\otimes V_{R_n}$ with prescribed
transformation properties under the complex gauge transformations
(gauge covariant). This infinite-dimensional definition is natural
in both the CS setting (where it corresponds to ``quantize first,
impose gauge conditions next'' approach) and in the WZW theory (it
produces there the solutions of the current algebra Ward identities).
Heuristically, our approach gives sections of a $\otimes
V_{R_n}$-vector bundle over the space of orbits of $G$-connections
under the complex gauge transformations which coincides
with the moduli space of holomorphic $G^{\NC}$-bundles on $\Sigma$.

The rigorous problem addressed is to describe the spaces of states
defined as above, together with the KZ connection,
in finite-dimensional terms and as explicitly as possible.
Another open (even at genus zero) mathematical problem which,
however, we shall not discuss below, is to prove that the KZ
connection is metric \cite{G}\cite{Gawedzki}\cite{FalGK}.
E. Verlinde, using the expected
factorization properties of the WZW conformal blocks together
with their modular properties has given a formula for the
dimension of the FS bundles \cite{Verl}\footnote{\cite{Verl}
contains the formula for the case with up to three insertions
on the sphere and for the arbitrary genus but no insertions;
its extension to the general case may be found in \cite{MS}}.
This formula has been rigorously proven in some special cases
and from various starting points\footnote{see
\cite{TsKa}\cite{TsKaUe} for the earlier related work}.

In \cite{AS} a proof for the $SU(2)$ case with no insertions
but in general topology was given within the moduli space approach.

Recently, G. Faltings has announced a proof of the general result
\cite{Falt}.

In \cite{GK}, a reduction
of the infinite-dimensional formulation to finite-dimensional
setup was achieved for $G=SU(2)$ and genus zero. The spaces of
spherical CS states were realized as subspaces of $Inv
(\otimes V_{R_n})$ composed of the $G$-invariant tensors satisfying
explicit conditions stable under the KZ connection.
The factorization properties of these spaces implying
the Verlinde formula for their dimensions where proven.
The present paper is, essentially,
an extension of this work to the genus 1 case. We realize
the toroidal CS states as theta-functions satisfying explicit
conditions. The flat (projective) connection allowing comparing
the states for different insertion points and different complex
structures on the torus is also constructed. The factorization
properties of the spaces of toroidal states when two insertion points
coincide allow us to prove the Verlinde formula modulo the assumption
that there are no states if one of the inserted spins is larger
than $k/2$. The latter fact was proven in the spherical case in
\cite{GK}. At genus one, it should follow by the factorization when
the torus is pinched to the sphere. We postpone the study of this
factorization, more difficult technically, to a future work.
\vskip .3cm

The organization of the paper is as follows. In Section 2 we define
precisely the theory and describe our picture of the Chern-Simons
states as given by gauge covariant holomorphic functionals on the space
of $G$-connections.
In Section 3 we describe a
stratification of the space of $SU(2)$-connections
($\cong$ the space of holomorphic $SL(2,\NC)$-bundles).
It is a refinement of the stratification given in \cite{AB}.
We need a finer decomposition of the semistable stratum
of \cite{AB} which then provides the main tool in the reduction
of the infinite-dimensional formulation to a finite-dimensional one.
The reduction associating to every state a set of theta-functions
is performed in Section 4. In Section 5, we construct the toroidal version
of the KZ connection on the FS bundle with fibers composed of
the toroidal CS states and the base given by the moduli space
of punctured complex tori. In Section 6, we
extend the bundle and the connection to the component of the boundary
of the base corresponding to two punctures coming together,
show the factorization of the fibers there
and prove the Verlinde formula for the dimension of the space of
states (all that modulo the assumption that there are no states with
spins $>k/2$). Finally in Section 7 we discuss our
conclusions and foresee future lines of research. Several
appendices collect proofs and properties that are used in the paper.

\nsection{\hspace{-.7cm}.\ \ Chern-Simons theory}

We shall study the CS
theory on a three-dimensional manifold
$\CM=\Sigma\times\NR$ where $\Sigma$ is a
compact Riemann surface without boundary and $\NR$
plays the role of time. The action of the theory
at level $k\in\NN$ and with Lie group $G=SU(n)$ is given by
\qq
S={_k\over^{4\pi}}\int_{\CM}\tr\s(BdB+{_2\over^3}B^3)
\qqq
where $B$ is the Yang-Mills connection on the trivial
$G$-bundle over $\CM$ i.e. an ${\gl su}(n)$-valued $1$-form
on $\CM$.
We shall also consider the insertion of $N$ disjoint Wilson lines
$P_n\times\NR$ carrying irreducible representations $R_n$ of $G$.
\vs 0.2cm

In the temporal gauge $B_0=0$, the phase space
of the theory is the set $\CA$ of two-dimensional connections $A$ on
the trivial
$G$-bundle over $\Sigma$ with symplectic
form ${k\over4\pi}\int\tr\s(\delta A)^2$.
Due to the remaining gauge freedom, this phase space
has to be constrained further by imposing an appropriate flatness
condition on connections $A$ \cite{Witten},\cite{EMSS} (see below).
\vs 0.2 cm

To quantize the theory, one may use the complex structure on $\CA$
induced by the complex structure on the Riemann surface:
the splitting $\CA\ni A=A_{\bar z} d\bar z+A_z dz$ allows to identify
$\CA$ with the complex space $\CA^{01}$ of ${\gl sl}(n,\NC)$-valued
$(0,1)$-forms on $\Sigma$. The quantum states in the holomorphic
quantization
{\it\`{a} la} Bargmann of the
theory are given
(see \cite{Gaw}\cite{EMSS}\cite{G}\cite{Gawedzki}\cite{GK})
by holomorphic functionals $\Psi$ of $A^{01}=
A_{\bar z}d\bar z\in\CA^{01}$
with values in $\otimes_{n=1}^N V_{R_n}$ where $V_{R}$ denotes the
vector
space of representation $R$.
States $\Psi$ must also satisfy the quantum flatness condition
\qq
(F^a(z)-\sum_{n=1}^N {_{2\pi}\over^k}\s\delta(z-z_n)\s
t^a_{(n)})\s\s\Psi(A^{01})=0
\label{gauss}
\qqq
where generators $t^a$ of ${\gl su}(n)$ are normalized
so that $\tr\s(t^at^b)={_1\over^2}\delta^{ab}$,
$\s t^a_{(n)}$ stands for the action of $t^a$ on the $n$-th factor of
$\otimes_n V_{R_n}$ and
\qq
F=\sum_aF^a t^a=\p_z A_{\bar z}
+\p_{\bar z}{_\pi\over^k}{\delta\over\delta A_{\bar z}}
-[{_\pi\over^k}{\delta\over\delta A_{\bar z}}, A_{\bar z}]
\qqq
is the curvature of $A=\sum_a A^a t^a$ with $A^a_z$ replaced by
its quantum version $-{_\pi\over^k}{\delta\over\delta A^a_{\bar z}}$.
\vskip 0.2cm

To introduce holomorphic functionals we need a differentiable
structure in
the space $\CA^{01}$ of smooth $(0,1)$-connections. We shall consider
this space
with the $C^\infty$ topology. By definition, holomorphic functionals
on it are $C^\infty$-smooth in the Fr\'echet  sense \cite{Ham} with
the
complex-linear derivative.
\vs 0.2cm

The flatness condition (\ref{gauss}) may be integrated in the
following
sense. Let $\CG^\NC$ denote the space of complex gauge
transformations i.e.
of smooth maps $h:\Sigma\rightarrow SL(n,\NC)$ which act on
$\CA^{01}$
by
\qq
A^{01}\mapsto\s{}^h{\hspace{-1mm}}A^{01}\equiv hA^{01}h^{-1}+h\bar\p
h^{-1}
\qqq
with $\bar\p=d\bar z \p_{\bar z}$. Then (\ref{gauss})
is equivalent to demanding that
\qq
\Psi({}^h{\hspace{-1mm}}A^{01})=\ee^{kS_{W\hspace{-0.05 cm}ZW}(h^{-
1},A^{01})}
\prod_n h(z_n)_{(n)}\s\Psi(A^{01}),
\label{blocks}
\qqq
for every $h\in\CG^\NC$. $S_{W\hspace{-0.05 cm}ZW}(h,A^{01})$ is
the action of the
Wess-Zumino-Witten (WZW)
model \cite{WittenBos} coupled
to the $(0,1)$-component of the gauge field.
Explicitly,
\qq
\hspace{-0.16cm}S_{W\hspace{-0.05 cm}ZW}(h,A^{01})=-{_i\over^{4\pi}}
\int_\Sigma \tr\s(h^{-1}\p h)(h^{-1}\bar\p h)
-{_i\over^{12\pi}}\int_\Sigma d^{-1}\tr\s(h^{-1}d h)^3
+{_i\over^{2\pi}}\int_\Sigma \tr\s(h\p h^{-1})A^{01}
\qqq
and is defined modulo $2\pi i$.
The cocycle property of the action
\qq
S_{W\hspace{-0.05 cm}ZW}(hh',{}^h{\hspace{-1mm}}A^{01})=
S_{W\hspace{-0.05 cm}ZW}(h',A^{01})-S_{W\hspace{-0.05 cm}ZW}(h^{-
1},A^{01})
\qqq
makes (\ref{blocks}) consistent under the product of gauge
transformations.
\vskip 0.2cm

If we formally define WZW Green functions by
the functional integral
\qq
\langle\Phi_1,\dots,\Phi_N\rangle_{1,A^{01}}
=\int\otimes_n g(z_n)_{(n)}\s\s\ee^{-kS_{W\hspace{-0.05
cm}ZW}(g,A^{01})}\prod_z dg(z)
\qqq
(here $g_{(n)}$ stands for the matrix representing $g$ in
representation $R_n$)
then, from the transformation of the action, one derives the Ward
identity
\qq
\langle\Phi_1,\dots,\Phi_N\rangle_{1,\s{}^h{\hspace{-1mm}}A^{01}}
=
\ee^{kS_{W\hspace{-0.05 cm}ZW}(h^{-1},\s A^{01})}
\prod_nh(z_n)_{(n)}\s\s
\langle\Phi_1,\dots,\Phi_N\rangle_{1,\s A^{01}}\label{Ward}
\qqq
which makes clear the relation between the space of states
of the Chern-Simons theory and the solutions of the chiral Ward
identity (\ref{Ward}) of the WZW model.
\vskip 0.2cm

{}From eq. (\ref{blocks}) we see that $\Psi$ will be determined
once we know its value at a point of each $\CG^\NC$ orbit.
However space $\CA^{01}/\CG^\NC$ is, in general, not a manifold
and the smoothness of $\Psi$ is not given for
free. In the following section we shall study
the relative positions and codimensions of the
$\CG^\NC$ orbits as an introductory step to finding the
holomorphic functionals $\Psi$ with transformation rule (\ref{blocks}).
\vskip .5cm

\nsection{\hspace{-.7cm}.\ \ Stratification of the space of connections}
\vskip 0.3cm

In this section we shall describe a
stratification of $\CA^{01}$ into submanifolds of finite
codimension invariant under the action of the gauge group.
\vskip 0.2 cm

First note that the space $\CA^{01}$ may be identified with the space
of holomorphic $SL(n,\NC)$ bundles (more exactly, of the
structures of a holomorphic $SL(n,\NC)$ vector bundle in the
trivial bundle $\Sigma\times\NC^n$). For $A^{01}\in\CA^{01}$,
$\bar\partial+A^{01}$ is the holomorphic derivative of the sections
of the corresponding holomorphic bundle. Isomorphic bundles
correspond
to gauge-related forms so that the space of orbits $\CA^{01}/\CG^\NC$
may be identified with the space of isomorphism classes of
holomorphic
$SL(n,\NC)$ vector bundles. Below, we shall use the bundle language
whenever it is more convenient.
\vskip 0.2 cm

We recall from \cite{AB} some definitions and results about the
holomorphic
vector bundles. Let $E$ be such a bundle, $n(E)$ its rank and $k(E)$
its Chern class ($k(E)=0$ if the bundle has structure group
$SL(n,\NC)$ ). The ratio
$\mu(E)=k(E)/n(E)$ is called the {\it slope} of $E$. $E$ is called
semistable
(stable) if for every proper holomorphic vector subbundle $D\subset
E$,
$\mu(D)\leq\mu(E)\ (\mu(D)<\mu(E))$.
Every holomorphic bundle $E$ has a canonical
filtration $$0=E_0\subset E_1\subset\cdots\subset E_r=E$$ such that
$D_i=E_i/E_{i-1}$ is semistable and
\qq
\mu(D_i)<\mu(D_{i-1}),\ i=1,\cdots,r.\label{filt}
\qqq
The ranks $n_i=n(D_i)$ and Chern classes $k_i=k(D_i)$, subject to
$\sum_i n_i = n(E)=n $ , $\sum_i k_i = 0$ and (\ref{filt}), determine
the stability type of $E$. Of course, if $E$ is semistable, $r=1$.
\vskip 0.2 cm

With the use of these data, $\CA^{01}$ is decomposed into submanifolds
 $\CA^{01}_\lambda$
formed by all bundles of type $\lambda=((n_i,k_i),\ i=1,\cdots,r)$.
As the type of a bundle is canonical, $\CA^{01}_\lambda$
is a union of $\CG^\NC$-orbits (isomorphism classes).  If $\CL$
denotes the set
 of all
possible types $\lambda$ then
\qq
\CA^{01}=\bigcup_{\lambda\in\CL}\CA^{01}_\lambda .\label{strat}
\qqq
One may partially order $\CL$ by setting $\lambda\preceq\mu$
if $\CP(\lambda)\subset\CP(\mu)$ where $\CP(\lambda)$ is the region
in the
two-dimensional plane between the horizontal axis and the convex
polygon
whose edges are vectors $(n_i,k_i)$ taken in order. The decomposition
(stratification) (\ref{strat}) has the following property \cite{AB}:

\qq
{\rm For\ every}\ I\subset\CL\ ,\ \
\CS_I=\bigcup_{\mu\in I}\bigcup_{\lambda\preceq\mu}
\CA^{01}_\lambda\ \ {\rm is\ an\ open\ submanifold\ of\ }\CA^{01}
\label{str1}
\qqq
\vskip 0.16 cm

\no The codimension of $\CA^{01}_\lambda$ is
\qq
c_\lambda=\sum_{r\geq i>j\geq 0}[(n_ik_j-n_jk_i)+ n_in_j(g-1)]
\label{cod}
\qqq
where $g$ is the genus of $\Sigma$.  The stratum $\CA^{01}_{ss}$ of
semistable bundles, associated to the minimal element
$\lambda_{ss}=((n,k))$
of $\CL$, is the only one of codimension zero and is an open dense
subset of $\CA^{01}$.
For the genus $g=1$ case under consideration in this paper,
we shall need to refine the above stratification, preserving still
property (\ref{str1}). This will be done
by decomposing the main stratum $\CA^{01}_{ss}$ into other
strata.
\vskip 0.2 cm

{}From the Weil theorem \cite{Gunn}, we know that each isomorphism
class of semistable holomorphic
vector bundles of vanishing Chern class (on a general
Riemann surface) has a flat
representative (for $g=1$ also the converse is true).
In other words, for each
semistable $A^{01}$ there exists an ${\gl sl}(n,\NC)$-valued
$(1,0)$-form $A^{10}$ such
that $A=A^{10}+A^{01}$ is flat
i.e. $F(A)=dA+A\wedge A=0$.
%(on $\CU_\alpha,\ A$ may be taken equal to
%$h_\alpha dh_\beta^{-1}$). Then, to analyze the space of
%orbits in $\CA^{01}_{ss}$ it is
%enough to consider flat bundles with constant transition
%functions $g_{\alpha\beta}=h_\alpha^{-1}h_\beta$.
To each of these bundles
(and to the choice of $A^{10}$) there corresponds a homomorphism
$$\rho: \pi_1(\Sigma)\longrightarrow SL(n,\NC)$$ given by the
holonomy
of the flat connection $A$:
\qq
A=hdh^{-1}\label{gauge}
\qqq
where the $SL(n,\NC)$-valued map $h$ is defined on the universal
covering
of $\Sigma$ and
\qq
h(a\xi)^{-1}=\rho(a)h(\xi)^{-1} {\rm\ \ for\ \ }a\in\pi_1(\Sigma).
\qqq
If the homomorhisms $\rho$ are related by
conjugation, then they come from isomorphic bundles (since the
corresponding flat connections are related by an $SL(n,\NC)$-valued
gauge transformation). In the toroidal case, i.e. when
$\Sigma=\NC/(2\pi\NZ+2\pi\tau\NZ)\ \ {\rm for}\ \
\tau\equiv\tau_1+i\tau_2,
\ \tau_2>0$,
we have a specially simple
situation as the fundamental group of $\Sigma$ is generated by two
commuting elements. Then the representations $\rho$ of the
fundamental
group are given by pairs of
commuting matrices $\rho_1, \rho_\tau\in SL(n,\NC)$.
\vskip 0.2 cm

For the rank $n=2$ to which we shall limit ourselves from now on,
 we can distinguish three classes of orbits
of equivalent semistable
bundles according to the behavior of $\rho$ for (one of) their flat
representatives
\vskip .4cm

\noindent {\bf Case 1.} One of the matrices is diagonalizable, and
different from $\pm I$, then both are simultaneously diagonalizable.
\vskip .2 cm

In this case, we may diagonalize by conjugation both matrices and
take:
$$\rho_\mu=\pmatrix{e^{\lambda_\mu}&0\cr0&
e^{-\lambda_\mu}}\equiv e^{\lambda_\mu\sigma_3}\ ,\qquad\mu=1,\tau.$$
We may still use the freedom of choice of the flat representative
for given bundle by replacing the map $h$ of (\ref{gauge}) by
$h\tilde{h}$ where $$\tilde{h}=e^{(2\pi)^{-1}z\lambda_1\sigma_3}.$$
This does not change $A^{01}$ but changes the transition matrices to
$$\rho_1=I,\ \ \rho_\tau=e^{2\pi i u\sigma_3}$$ where $2\pi iu=
\lambda_\tau-\tau\lambda_1$. One can take
\qq
A_u=A^{10}_u+A^{01}_u=u\sigma_3{(dz-d\bar z)}/{2\tau_2}\label{flat}
\qqq
as the associated flat connection.
\vskip 0.2 cm

There are gauge relations between chiral connections $A^{01}_u$. Let
\qq
h_{v}=\exp[{{_{v\bar z-\bar
 vz}}\over{^{2\tau_2}}}\sigma_3]\label{gautr}\label{h}
\qqq
be a map from the complex plane to $SL(2,\NC)$. For $v=m+\tau
n\in\NZ+\tau\NZ$,
$h_v$ defines a gauge transformation on the torus which shifts
$u$ of $A^{01}_u$ by $m+\tau n$. Also, the constant gauge
transformations with
values in the normalizer of the Cartan subgroup map $A^{01}_u$ to
itself or to
$A^{01}_{-u}$.
A direct check shows that for $u\not\in(\NZ+\tau\NZ)/2$ these
generate the only
gauge transformations relating $A^{01}_u$'s. To generate all the
gauge
 transformations
relating $A_{u=0}^{01}$ to other $A^{01}_u$'s, one has to add the
arbitrary
 constant
gauge automorphisms $g_0$ of $A_0^{01}$ to $h_v$'s with
$v\in\NZ+\tau\NZ$. To
 understand the
gauge relations of an arbitrary $A^{01}_u$, $u\in(\NZ+\tau\NZ)/2$, to
other
 $A^{01}_u$'s,
notice that for $v\in(\NZ+\tau\NZ)/2$, although
$h_v\not\in\CG^\NC$ if $v\not\in\NZ+\tau\NZ$ as it is
multivalued on the torus, it still defines an automorphism
$A^{01}\mapsto\s{}^{h_v}A^{01}$
of $\CA^{01}$ since $h_v$ is multiplied by $\pm I$ when
one goes around the torus' cycles. The action of such $h_v$ still
shifts
$u$ of $A^{01}_u$ by $v\in(\NZ+\tau \NZ)/2$ and we may use it to
infer all gauge automorphism connecting any $A_u^{01}$ with
$u\in(\NZ+\tau\NZ)/2$
to any other $A_u$'s. They are generated by the automorphisms
$h_ug_0h_u^{-1}$
of $A_u^{01}$ ($u\in(\NZ+\tau\NZ)/2$) and $h_v$'s with
$v\in\NZ+\tau\NZ$.
\vskip 0.2 cm

It follows that we may
split the union of $\CG^\NC$-orbits of chiral connections $A^{01}_u$
into five disjoint sets. $\CA^{01}_0$ will contain the orbits of
$A^{01}_u$
with $u\not\in(\NZ+\tau\NZ)/2$.
The orbits of $A^{01}_u$'s with $u\in\alpha+\NZ+\tau\NZ$, $\alpha=
0, 1/2, \tau/2, (1+\tau)/2$ will form the sets
$\CA^{01}_{(\alpha,1)}$.
\vskip .5cm

\noindent {\bf Case 2. } One of the matrices is not diagonalizable.
\vskip 0.3 cm

Then we can always, by conjugation, reduce the two matrices to the
form:
$$\rho_\mu=\pm\pmatrix{1&a_\mu\cr0& 1}\equiv\pm e^{a_\mu\sigma_+},
\qquad\mu=1,\ \tau.$$ If
$a_\tau\not=\tau a_1$ then using $$\tilde{h}=e^{(2\pi)^{-
1}za_1\sigma_+}$$
as in Case 1 and with a further conjugation, we may equivalently
obtain
\qq
\rho_1=\pm I,\qquad
\rho_\tau=\pm \pmatrix{1&2\pi i\cr0&1}
\qqq
where all four possibilities of combinations of signs
$\pm$ are allowed. As the corresponding flat connections we may take
$$\hat A_\alpha=h_\alpha(\sigma_+(dz-d\bar
z)/(2\tau_2))h_\alpha^{-1}+h_\alpha dh_\alpha^{-1}$$
where, for $\alpha=0,1/2,\tau/2,(1+\tau)/2$ corresponding
to signs $(+,+)$,$(+,-)$,$(-,+)$,$(-,-)$,
$h_\alpha$ are given by (\ref{h}).
\vskip 0.2 cm

The orbits of
$$\hat A_\alpha^{01}=-(\alpha\sigma_3+{e}^{{\alpha\bar z-
\bar\alpha z\over\tau_2}}\sigma_+)d\bar z/(2\tau_2)$$
will form the subsets $\CA^{01}_{(\alpha,0)}$ of $\CA^{01}$.
Bundles with $a_\tau=\tau a_1$, excluded from
$\CA^{01}_{(\alpha,0)}$, are
equivalent to those with transition functions $\pm I$, and they are
elements of the sets $\CA^{01}_{(\alpha,1)}$ considered above.
\vskip 0.3 cm

{}From the remark that the equivalence of holonomies $\rho$ implies
that
 $A^{01}$'s
are in the same orbit it should be clear that the sets $\CA^{01}_0$,
$\CA^{01}_{(\alpha,i)}$
cover the semistable stratum $\CA^{01}_{ss}$ (they produce all
possible
nonequivalent
holonomies). It may be less clear that
they are all disjoint since they were singled out according to the
properties
of their (non-unique) flat representatives. Notice however that if
$A=A^{10}+A^{01}$ and $A'=A'^{10}+A^{01}$ are two flat
representatives
of the same $A^{01}$ then $\omega=A^{10}-A'^{10}$ satisfies
$$\bar\partial\omega+A^{01}\wedge\omega+\omega\wedge A^{01}=0.$$
A direct check shows that for $A^{01}=\hat A^{01}_\alpha$ the
solutions
are of the form $\omega_\alpha=h_\alpha(a\sigma_+dz)h_\alpha^{-1}$
with complex
 $a$
and all lead to the same behavior of the holonomy matrices. This
demonstrates
that sets $\CA^{01}_{(\alpha,0)}$ are mutually disjoint and disjoint
from the
union of orbits of $A^{01}_u$'s forming (disjoint) sets $\CA^{01}_0$
and
$\CA^{01}_{(\alpha,1)}$.
\vskip 0.5 cm

Now we may assert the main result of this section:

\vskip .4cm
Take the set of indices $\hat\CL=(\CL\setminus\{\lambda_{ss}\})\cup
\{0\}\cup\{(\alpha,0),(\alpha,1)\s|\s\alpha=0,1/2,\tau/2,(1+\tau)/2\}$.
Extend the order relation defined in $\CL$ to $\hat\CL$ by
$$0\preceq(\alpha,0)\preceq(\alpha,1)\preceq\lambda\qquad\forall\lambda\in
\CL\setminus\{\lambda_{ss}\}\ \ {\rm and}\ \ \alpha=0, {1/2},
{\tau/2}, {(1+\tau)/2}.$$
\vskip 0.2 cm

Then:
\vskip 0.2 cm

\noindent {a)} $\CA^{01}_{\hat\lambda}\ {\rm for}\
\hat\lambda\in\hat\CL$ are connected,
mutually disjoint submanifolds of $\CA^{01}$. They are invariant
under the
action of $\CG^\NC$. $\displaystyle
\CA^{01}=\bigcup_{\hat\lambda\in\hat\CL}\CA^{01}_{\hat\lambda}$.

\noindent {b)} For every $\hat I\subset\hat\CL$
$$\CS_{\hat I}=\bigcup_{\hat\mu\in \hat I}\s
\bigcup_{\hat\lambda\preceq\hat\mu}
\CA^{01}_{\hat\lambda}\qquad {\rm is\ an\ open\ submanifold\ of\
}\CA^{01}.$$
Consequently, $\CA^{01}_{\hat\lambda}$ is a closed submanifold of
$\CS_{{\hat I}\cup\{\hat{\lambda}\}}$ for
any minimal element $\hat\lambda$ of $\hat\CL\setminus\hat I$.
\vskip 0.3 cm

\noindent{c)} Codimensions $c_{\hat\lambda}$ of
$\CA^{01}_{\hat\lambda}$
are: $c_0=0$ ($\CA^{01}_0$ is an open dense subset of $\CA^{01}$),
 $c_{(\alpha,0)}=1$,
$c_{(\alpha,1)}=3$ and $c_{\hat\lambda}$ for
$\hat\lambda\in\CL\setminus\{\lambda_{ss}\}$ is as in (\ref{cod}).

\vskip .5cm

\no{\bf Proof}.
\vskip 0.2 cm

Let $\CG^\NC_0$ be the group of gauge transformations
which are the identity at the origin and $\CC\subset\CA^{01}$
the set of constant connections. The elements of the latter may be
written as
\qq
A^{01}_M=-\frac{_1}{^2}\tau_2^{-1}Md\bar z,\qquad M\in{\gl
sl}(2,\NC).\label{const}
\qqq
Fix a real number $0<\epsilon<1/4$ and take the open set
$U_0\subset\CC$
of constant connections associated to matrices $M\in\hbox{\gl
sl}(2,\NC)$ such
that
\qq
\det M= (a+b\tau)^2\ \ \ \ \ {\rm with }\ \
a,b\in(-1/2+\epsilon, 1/2-\epsilon)\subset
\NR.\label{bound}
\qqq

\no Consider the map
\qq
P_0:\CG_0^\NC\times U_0&\longrightarrow& \CA^{01}_{ss}\cr
(g,A_M^{01})&\longmapsto& {}^g A_M^{01}.\label{map}
\qqq
It follows then by the Nash-Moser (inverse function)
Theorem \cite{Ham} that
$P_0$ is a smooth
diffeomorphism onto its (open) image, see Appendix A.
\vskip 0.2 cm

Now take
\qq
\hbox to 1cm{$V_0$\hfill}&=&\{A^{01}_M\in U_0\s|\s\det(M)\not=0\}\s,\cr
\hbox to 1cm{$V_{(0,0)}$\hfill}&=&\{A^{01}_M\in U_0\s|\s\det(M)=0 {\rm\
and\ } M\not=0\}\s,\cr
\hbox to 1cm{$V_{(0,1)}$\hfill}&=&\{0\}\subset U_0\s.
\qqq
Then $\CA^{01}_{(0,i)}=P_0(\CG_0^\NC\times V_{(0,i)})$ and $P_0(V_0)$
is an open
 subset
of $\CA^{01}_0$. It follows that $\CA^{01}_{(0,0)}$
and $\CA^{01}_{(0,1)}$ have codimensions $1$ and $3$, respectively.
\vskip 0.1 cm

For the three other values of $\alpha$, instead
of constant connections, we take $U_\alpha={}^{h_\alpha} U_0$ with
$h_\alpha$ given by (\ref{h}) and
\qq
P_\alpha:\CG_0^\NC\times U_\alpha&\longrightarrow& \CA^{01}\cr
(g,A^{01})&\longmapsto& {}^g A^{01}.\label{alpha}
\qqq
It is clear that $P_\alpha$ is also a smooth diffeomorphism
onto its image.
Then, for $V_\alpha={}^{h_\alpha}\hspace{-0.05 cm}
V_0$, we have $\CA^{01}_0=\bigcup_\alpha
P_\alpha(\CG_0^\NC\times V_\alpha)$ and so it is an open manifold
dense in
$\CA^{01}_{ss}$ and, consequently, also in $\CA^{01}$. On the
other hand $\CA^{01}_{(\alpha,i)}={}^{h_\alpha}\hspace{-0.05 cm}
\CA^{01}_{(0,i)}, \ i=0,1$. From
this, and from the properties of the initial stratification
(\ref{str1}), \ a),
b) and c) follow immediately.

\vskip 1cm

\nsection{\hspace{-.7cm}.\ \ Space of states}
\vskip 0.3 cm

With the detailed knowledge of the space $\CA^{01}$ of the chiral
connections provided by the preceding Section, we are
in a position to describe more effectively the Chern-Simons states
defined
as the holomorphic functionals $\Psi$ on
$\CA^{01}$ taking values in $\bigotimes_n\hspace{-0.05 cm} V_{j_n}$
($j_n$ are spins of the Wilson lines)
and transforming according
to (\ref{blocks}) under the (complex) gauge transformations. By
considering
such a state on the one-parameter family of connections $A^{01}_u$ of
(\ref{flat}),
we obtain a holomorphic map $\gamma:\NC\longrightarrow
\bigotimes_n\hspace{-0.05 cm}V_{j_n}$ given by
\qq
\Psi(A_u^{01})\equiv e^{\pi k u^2/\tau_2}
\prod_n (e^{\sigma_3(\bar z_n-z_n)u/(2\tau_2)})_{(n)}
\ \gamma(u)\hspace{.06cm}.\label{blocks1}
\qqq
The gauge transformations by $h_{m+\tau n}$ (see (\ref{h})) and by
elements of the normalizer of the Cartan subgroup give rise to the
conditions
\qq
\gamma(u+1)\hspace{0.04 cm}&=&\gamma(u),\aeqno a\cr
\ &\cr
\gamma(u+\tau)&=&e^{-2\pi i k  (\tau +2 u)}\hspace{.06 cm}
\prod_{n} (e^{i z_n\sigma_3})_{(n)}\ \gamma(u),\alabel{cond}b\cr
\ &\cr
0\ \ \ \ \ \ &=&\sum_{n} (\sigma_3)_{(n)}\ \gamma(u)
\aeqno c\cr
\ &\cr
\gamma(-u)\ \ \ &=&\prod_{n} (\Omega)_{(n)}\ \gamma(u)\aeqno d\cr
\qqq
where $\Omega$ is the generator of the Weyl group of $SL(2,\NC)$.
Conversely, a holomorphic
map $\gamma$ on $\NC$ satisfying the above conditions
defines by (\ref{blocks1}) and (\ref{blocks}) a unique functional
$\Psi$ on the open dense stratum $\CA^{01}_0$ satisfying there
(\ref{blocks}).
We shall see that such $\Psi$ is holomorphic on $\CA^{01}_0$
but need not extend to the whole of $\CA^{01}$ (if it does then
the extension is unique and satisfies (\ref{blocks}) everywhere).
We shall determine below the necessary and sufficient conditions for
such an extension to exist. The results of the last Section will be
crucial here.
\vskip 0.2 cm

In the polynomial realization of the spaces
$V_j$ of spin $j$ representations of $SL(2,\NC)$, elements of
$V_j$
are polynomials of degree at most $2j$ with the action of the
group by $$\pmatrix {a&b\cr c&d}^{-1} P(v)= (cv+d)^{2j}\hspace{0.06
cm}
P\Bigl({av+b\over cv+d}\Bigr).$$ In this realization, $\gamma(u)$
is a polynomial in variables ${\underline v}\equiv(v_n)$ associated
to spaces $(V_{j_n})$.
Using (\ref{cond},c) we can see that $\gamma(u)$ is
homogeneous of degree $J\equiv\sum_nj_n$ in variables $\underline v$.
In
particular, this implies that $J$ has to
be an integer.  We may write
\qq
\gamma({u})=
\sum_{\underline m}f_{\underline
m}({u})\prod_nv_n^{j_n+m_n}\label{states}
\qqq
where the sum is over the
N-tuples ${\underline m}\equiv(m_n)$  such that $0\leq j_n-
|m_n|\in\NZ$ and $\sum_n
m_n=0$.
{}From (\ref{cond},d) we infer that
\qq
f_{\underline m}({u})=(-1)^J f_{-\underline m}(-{u}).\non
\qqq
Finally, from (\ref{cond},a-b),
\qq
f_{\underline m}({u})= \vartheta({u}-(4\pi k)^{-1}\sum_n
z_nm_n)\label{comp}
\qqq
with $\vartheta$ an ${\underline m}$-dependent theta-function
of degree $2k$
satisfying
\qq
\vartheta({u}+1)=\vartheta({u})\ ,\qquad
\vartheta({u}+\tau)=e^{-2\pi i
k(\tau+2{u})}\vartheta({u}).\label{theta}
\qqq
There are $2k$ independent solutions $\vartheta_p$ of eqs.
(\ref{theta})
given by
$$\vartheta_p(u)=\sum\limits_{n=-\infty}^\infty
q^{(p+2kn)^2/(4k)}\ee^{2\pi iu(p+2kn)}$$
where $p=0,1,...,2k-1$ and $q\equiv e^{2\pi i\tau}$.
Consequently, the spaces of analytic maps
$\gamma:\NC\rightarrow\otimes_nV_n$ satisfying (\ref{cond})
for fixed $\tau$ and $\underline z\equiv(z_n)$, $z_n\in\NC$, are
of finite, constant dimension and form a holomorphic vector
bundle $W_{\un j}$ over the space $\CT_N$ of $(\un z,\tau)$
with $z_n\not=z_{n'}\hs{0.07 cm}{\rm mod}\hs{0.04
cm}(2\pi,2\pi\tau)$.
\vskip 0.2 cm

The above description of states $\Psi$ defined on flat connections
$A^{01}_u$ has used a choice of complex coordinates $z_n$ whereas the
class
of connections $A^{01}_u$ depends only on the complex structure of
the torus. To exhibit the geometric character of $\gamma$'s
notice that different choices of $z_n$'s are intertwined by the
action
of the discrete group
$\Gamma_N=SL(2,\NZ)\hs{0.08 cm}{\dl n}\NZ^{2N}$
where $\left(\pmatrix{a&b\cr c&d},(p_n,r_n)\right)\in\Gamma_N$
acts by mapping $(\un z,\tau)$ to $(\un z',\tau')$
where
\qq
\tau'=(a\tau+b)/(c\tau+d),\ \ \ \
z_n'=(z_n+2\pi p_n+2\pi\tau r_n)/(c\tau+d)\label{mod}.
\qqq
The action of $\Gamma_N$ on
bundle $W_{\un j}$ which lifts the action
$(\un z,\tau)\mapsto(\un z',\tau')$ on the
base is given by the formula
\qq
\gamma'(u')=
e^{2\pi ikcu'u}\hs{0.06 cm}
\prod_n
\left(e^{-icz'_nu\sigma_3+2\pi ir_nu\sigma_3}\right)_{(n)}\
\gamma(u),
\qqq
where $u'=u/(c\tau+d)$.
$\gamma$'s related by this action describe
the same state $\Psi$ for different choices of complex
coordinates $z_n$ (notice that $A_u\equiv-u\sigma_3d\bar z/(2\tau_2)
=A'_{u'}\equiv-u'\sigma_3d\bar z'/(2\tau'_2)$). Dividing
bundle $W_{\un j}$ by the action of $\Gamma_N$, one obtains a
bundle $W_{\un j}/\Gamma_N$ over the moduli space
$\CT_N/\Gamma_N$ of tori with $N$ punctures.
\vskip 0.2 cm

Let us return to discussing which $\gamma$'s correspond to
global states $\Psi$.
By explicitly conjugating a constant
matrix $M\equiv\pmatrix{M_0&M_1\cr M_2&-M_0}
\in{\gl sl}(2,\NC)$ such that $\det(M)\not=0$ and $M_1\not=0$ to
a diagonal matrix $u\sigma_3$ and by
using (\ref{blocks1}) and (\ref{blocks}),
we obtain the following expression for
$\Psi(^{h_\alpha}\hspace{-0.11cm}A^{01}_M)$
(see (\ref{const}) and (\ref{h})):

\qq
&\displaystyle\prod_n
\left(e^{M(z_n-\bar z_n)/(2\tau_2)h_\alpha(z_n)}\right)_{(n)}\
\Psi(^{h_\alpha}\hspace{-0.11cm}A^{01}_M)\big((
v_n)\big)\cr
&\cr
&\displaystyle=e^{\pi k((u+\alpha)^2
+2(M_0-u)\bar\alpha)/\tau_2}\
\left(\prod_ne^{-(\alpha-\bar \alpha) z_nj_n/\tau_2}
(M_1+(M_0-u)v_n)^{2j_n}\right)\cr
&\cr
&\displaystyle\cdot\ (2M_1u)^{-J}\ \gamma(u+\alpha)
\left((e^{(\alpha-\bar\alpha)z_n/\tau_2}
(M_1+(M_0+u)v_n)/(M_1+(M_0-u)v_n))
\right)\label{bonc}
\qqq
\vskip .15 cm
\no where $u=\pm(M_0^2+M_1M_2)^{1/2}$ and properties (\ref{cond},a-d)
of $\gamma$ insure that the right hand side is independent of
the choice of the sign. This rather messy expression becomes
much simpler for $M(u)=\pmatrix{u&2\cr 0&-u}$:
\qq
&\displaystyle\prod_n\left(e^{M(u)
(z_n-\bar z_n)/(2\tau_2)}h_\alpha(z_n)\right)_{(n)}
\ \Psi(^{h_\alpha}\hspace{-0.11cm}A^{01}_{M(u)})
\big((v_n)\big)\cr
&\cr
&\displaystyle=e^{\pi k(u+\alpha)^2/\tau_2}\hspace{0.13 cm}
\big(\prod_n e^{-(\alpha-\bar \alpha)z_n j_n/\tau_2}\big)
\hs{.13cm}u^{-J}\hs{0.15 cm}
\gamma(u+\alpha)
\left((e^{(\alpha-\bar\alpha)z_n/\tau_2}
(1+uv_n))\right).
\label{alpha1}
\qqq
Notice the negative power $u^{-J}$ on the right hand side which
is entire in $u$ if and only if
\vskip 0.1 cm
\qq
\hat D^L_{{\un j},\alpha}\gamma\equiv
\partial_u^{l_0}\partial_{v_1}^{l_1}\cdots\partial_{v_N}^{l_N}
\gamma(u)({\underline v})\vert_
{u=\alpha\hfill\atop v_n=\exp[(\alpha-\bar\alpha)z_n/\tau_2]}=0
\label{reg1}
\qqq
\hspace*{1cm}for every $N+1$-tuple of non negative integers
$L\equiv(l_n)$ such that $\displaystyle|L|\equiv\sum_{n=0}^{N} l_n <
J$.
\vskip 0.55 cm

\no Conversely, it is not difficult to see that
this condition implies that $\Psi(^{h_\alpha}
\hspace{-0.11cm}A^{01}_M)$, as given by
(\ref{bonc}), is analytic on the set of $M$'s with $M_1\not=0$. The
last
requirement may be relaxed to $M\not=0$ by conjugating matrices $M$.
Thus condition (\ref{reg1}) is  necessary and sufficient
for $\Psi(^{h\alpha}\hspace{-0.06 cm}A^{01}_M)$ to
extend analytically to the codimension $1$
stratum of matrices $M\not=0$ with $det(M)=0$.
\vskip 0.2 cm

Now using maps $P_\alpha$ of (\ref{alpha}), we infer the
holomorphicity
of $\Psi$ on the open dense stratum $\CA^{01}_0$ and see that
it extends analytically to the codimension $1$ strata
$\CA^{01}_{(\alpha,0)}$
if and only if the conditions (\ref{reg1}) are satisfied. The
extension
to the other strata of $\CA^{01}$, of codimension $>1$, follows then
automatically by applying inductively the Hartogs Theorem. This is
there that one uses property b) of the
stratification of $\CA^{01}$ proven in Section 3.
\vskip 0.2 cm

\underline{Summarizing}:  a holomorphic map $\gamma:\NC
\longrightarrow\bigotimes_nV_{j_n}$ satisfying conditions
(\ref{cond},a-d)
defines by relations (\ref{blocks1}) and (\ref{blocks}) a (global)
Chern-Simons state if and only if it satisfies (\ref{reg1}) for
$\alpha=0,1/2,\tau/2,(1+\tau)/2$; besides, every Chern-Simons state
may be uniquely represented this way.
\vskip 0.3 cm

It is not difficult to see that condition (\ref{reg1}) is preserved
by the action of the discrete group $\Gamma_N$ (notice that
$A'^{01}_{M(u')}=
e^{(c\tau+d)^{1/2}\sigma_3}
A^{01}_{M(u)}e^{-(c\tau+d)^{1/2}\sigma_3}$\hs{0.06 cm}).
If the spaces of solutions of (\ref{reg1}) are of constant dimension,
they form a holomorphic (``fusion rules'')
subbundle $W^{fr}_{\un j}$ of
$W_{\un j}$ equivariant under the action of $\Gamma_N$ and
consequently
projecting to a bundle over the moduli space $\CT_N/\Gamma_N$.
\vskip 0.3 cm

As an illustration, let compute the spaces
of Chern-Simons states with zero and one insertions.
\vskip 0.3 cm

\no{\bf States with zero insertions.}
\vskip 0.2 cm

{}From (\ref{states}) and  (\ref{comp}) the zero-points states are
simply
the even theta-functions of degree $2k$ since for
$J=0$ conditions (\ref{reg1}) play no role.
The space of even theta-functions has dimension $k+1$. It is
spanned by the Kac-Moody characters
$\chi_{k,j}(\tau,\ee^{2\pi iu})$ with fixed $\tau$ and
$j=0,1/2,\dots,k/2$.
\vskip 0.3 cm

\no{\bf States with one insertion.}
\vskip 0.2 cm

In this case the states with one insertion of spin $j\in\NN$ are
\qq
\gamma(u)(v)=\vartheta(u)v^j,
\qqq
where $\vartheta$ is a theta-function of degree $2k$ with
\qq
&\vartheta(-u)=(-1)^j\vartheta(u)\s,\alabel{1p}a\cr\cr
&\partial_u^l\vartheta(u)|_{u=\alpha}=0&\qquad\hbox{for $l<j$ and
$\alpha=0$,$1\over 2$,$\tau\over 2$,$\tau+1\over 2$\s.}\hskip .5cm
\aeqno b\cr
\qqq
\vskip -\baselineskip
To study the dimension $\CD_j$ of the space of solutions of
(\ref{1p},a-b) we shall first focus on the case
of even spin, $j\in 2\NN$. In that case (\ref{1p},a) requires
even theta-functions for which:
\qq
&\ee^{-4 \pi i ku\textstyle{\alpha-\bar\alpha\over\tau-\bar\tau}}
\vartheta(u+\alpha)=
\ee^{4 \pi i ku\textstyle{\alpha-\bar\alpha\over\tau-\bar\tau}}
\vartheta(-u+\alpha)&\hbox{\hskip 1cm$\alpha=0$,
$1\over 2$,$\tau\over 2$,$\tau+1\over 2$.}\label{eveth}
\qqq
As a consequence of (\ref{eveth}), vanishing of
$\partial_u^l\vartheta(u)|_{u=\alpha}$ for $l\leq 2n\in 2\NN$ implies
also vanishing of the same expression for $l=2n+1$. So in
(\ref{1p},b),
we may consider only even values of $l$. Thus we are left with $2j$
linear conditions on the $k+1$ dimensional space of even
theta-functions
and we obtain the lower bound on the dimension of the space
of solutions:
\qq
\CD_j\geq k-2j+1\s.\label{lbound}
\qqq

To obtain an upper bound for $\CD_j\s$, \s we shall use the fact
that the sum of multiplicities of zeros of a theta-function
of degree $2k$
in any fundamental cell is $2k$. Now, if we have $\CD_j$ independent
even theta-functions with common zeros of total multiplicity $4j\leq
2k$,
by a linear combination of them we can obtain another
theta-function whose zeros have multiplicity at least $4j+2(\CD_j-1)$.
Since the latter number has to be $\leq 2k$, it follows that
$\CD_j\leq k-2j+1$
and consequently we obtain the expected result
\qq
\CD_j=\cases{
\matrix{k-2j+1 \hfill&&\hbox{for \ }j\leq k/2\s,\hfill\cr
0\hfill&&\hbox{otherwise\s.}\hfill}}\label{dim}
\qqq

The case of odd spin $j$ can be treated in a similar way. Now
we have $k-1$ independent odd theta-functions
and (\ref{1p},b) gives $2j-2$ independent conditions. Finally
we obtain the same expression (\ref{dim}) for the $\CD_j$.

\vskip 1cm

\nsection{\hspace{-.7cm}.\ \ Knizhnik-Zamolodchikov-Bernard equations}
\vskip 0.3 cm

In Section 2 we have described how the states of Chern-Simons theory
may be
identified with holomorphic factors of (euclidean) Green functions
for the WZW
theory given by a formal functional integral. Using the
functional integral
representation, one may deduce equations
that describe the behavior of the Green functions under changes of
the conformal
structure of the torus and of the insertion points. In the spherical
topology,
such relations are known as the Knizhnik-Zamolodchikov (KZ) equations
\cite{KnizhZamo}.
They were generalized to the toroidal case in \cite{Denis1} and to
general
Riemann surfaces in \cite{Denis2}.
They induce a connection which allows comparing spaces of
Chern-Simons states for
different punctured surfaces, see
\cite{Witten}\cite{Hitchin}\cite{Gawedzki}
\cite{AxelWitten}.
We have included a brief heuristic derivation of the
toroidal equations for the sake of completeness.

First, let us consider twisted toroidal Green functions of the WZW
theory
coupled to the $(0,1)$-component of the connection:
\qq
\langle \Phi_1( z_1)\cdots\Phi_N( z_N)\rangle_{\eta,A^{01}}=
\int  \otimes_n(g_1 g)( z_n)_{(n)}
\hspace{0.07 cm}\exp[ -k S_{W\hspace{-0.05
cm}ZW}(g_1g,A)]\hspace{0.11 cm}Dg\s.
\label{Green}
\qqq
Field $g$ in (\ref{Green}) is periodic i.e. $g(z+2\pi)=
g(z+2\pi\tau)=g(z)$
and  $g_1(z+2\pi)=g_1(z)$, $g_1(z+2\pi\tau)=\eta g_1(z)$ with
twist $\eta\in G^{\NC}$. Also $A^{01}(z+2\pi)=A^{01}(z)$,
$A^{01}(z+2\pi\tau)={\rm Ad}_\eta A^{01}(z)$.
The WZW action
for the twisted fields may be defined \cite{FGK} so that if
$g_1'(qz)=
\eta'g_1'(z)$, $A^{01}=g_1{g_1'}^{-1}\bar\partial
(g_1'{g_1}^{-1})$ then
\qq
S_{W\hspace{-0.05 cm}ZW}(g_1g,A^{01})=S_{W\hspace{-0.05
cm}ZW}(g_1'g)-
S_{W\hspace{-0.05 cm}ZW}(g_1')
+S_{W\hspace{-0.05 cm}ZW}(g_1,A^{01}).\label{acttwist}
\qqq
In particular, for $g'_1(z)=\ee^{\sigma_3(z-\bar z)u/(2\tau_2)}\equiv
\eta_u^{-1}g'_1(z+2\pi\tau)$ and $g_1=1$, we have
$$S_{W\hspace{-0.05 cm}ZW}(g,A^{01}_u)=S_{W\hspace{-0.05
cm}ZW}(g_1'g)
-\pi u^2/\tau_2$$
and we obtain the following relation between the untwisted and
twisted
Green functions:
\qq
\langle \Phi_1( z_1)\cdots\Phi_N( z_N)\rangle_{1,A^{01}}
= e^{\pi k u^2/\tau_2}
\prod_n (e^{\sigma_3(\bar z_n-z_n)u/(2\tau_2)})_{(n)}
\langle \Phi_1( z_1)\cdots\Phi_N( z_N)\rangle_{\eta_u,0}.
\label{cor}
\qqq
As we saw before, Chern-Simons states $\Psi(A^{01})$
have the same transformation properties under chiral gauge
transformation as
$\langle\cdots\rangle_{1,A^{01}}$. Comparing
(\ref{cor}) and (\ref{blocks1}), one infers that, similarly,
maps $\gamma(u)$ correspond to $\langle\cdots\rangle_{\eta_u,0}$
for $\eta_u= \exp(2\pi i u\sigma_3)$.

Eq. (\ref{acttwist}) implies the following Ward identity for the
Green functions:
\qq
\langle \Phi_1( z_1)\cdots\Phi_N( z_N)\rangle_{\eta,A^{01}}&=&
\ee^{k(S_{W\hspace{-0.05 cm}ZW}(g_1')-S_{W\hspace{-0.05
cm}ZW}(g_1,A^{01}))}
\cr
&&
\cdot\ \prod_n (g_1g_1'^{-1})(z_n)_{(n)}\hspace{0.08 cm}
\langle \Phi_1( z_1)\cdots\Phi_N( z_N)\rangle_{\eta',0}.
\label{corr}
\qqq
On the other hand, correlations $\langle\cdots\rangle_{\eta,A}$ are
generating functionals for Kac-Moody currents. For example,
\qq
\langle J^a(z) J^b(w)\Phi_1\cdots\Phi_N\rangle_{\eta,0}=
(2\pi i)^2 {\delta^2\over\delta (A^{01})^a(z)\delta (A^{01})^b(w)}
\langle\Phi_1\cdots\Phi_N\rangle_{\eta,A}\vert_{A=0}.
\qqq
where $A^{01}=\sum_a(A^{01})^at^a$. We shall normalize the Lie
algebra generators $t^a$ so that ${\rm tr}\hs{0.06
cm}t^at^b={_1\over^2}
\delta^{ab}$.
Identity (\ref{corr}) allows to compute
the Green functions
with two  current insertions if one solves
\qq
A^{01}&=&h^{-1}\bar\partial h,\alabel{eqA}a\cr
 h(z+2\pi)&=&h(z),\aeqno b\cr
 h(z+2\pi\tau)&=&\eta' h(z) \eta^{-1}
\aeqno c\cr
\qqq
\vskip-\baselineskip\no
for $h$ and $\eta'$ to the second order in $A^{01}$.  The latter may
be
expressed in terms of the fundamental solution $\omega(z,w|\eta)$
of the operator $\bar\partial$ with boundary condition
$$\omega(z+2\pi,w|\eta)=\omega(z,w|\eta),$$
$$\omega(z+2\pi\tau,w|\eta)={\rm Ad}_\eta \omega(z,w|\eta)-i\s.$$
More explicitly,
\qq
\omega(z,w|\eta)={i\ee^{iw}\over\ee^{iz}-\ee^{iw}}+
\sum_{r=1}^\infty\Bigl(
{iq^r\over Ad_{\eta^{-1}}-q^r}\ee^{ir(w-z)}-
{iq^r\over Ad_{\eta}-q^r}\ee^{ir(z-w)}\Bigr).\label{omega}
\qqq
where $\s q\equiv\ee^{2\pi i\tau}\s.\s$
Now solutions of (\ref{eqA},a-c) are given by
\qq
\eta'\eta^{-1}&=& \exp\Bigl
[{_1\over^{2\pi}}\int dw A^{01}(w)
\cr&&\hskip .7cm+{_i\over^{8\pi^2}}\int\int
[\omega(z,w|\eta)dwA^{01}(w),dzA^{01}(z)]+ \CO(A^3)\Bigr],
\aeqno a\cr
\cr
h(z)&=&\exp\Bigl[{{_i}\over^{4\pi}}\int
\Bigl(\omega(z,w|\eta)+\omega(z,w|\eta')\Bigr)dwA^{01}(w)
\cr&&\hskip .7cm-{_{1}\over^{8\pi^2}}\int\int
\omega(z,y|\eta)[\omega(y,w|\eta)dwA^{01}(w),dyA^{01}(y)]+
\CO(A^3)\Bigr].\aeqno b\cr.
\qqq
Using those expressions on the right hand side of (\ref{corr}) one
obtains for the Green functions with two current insertions:
\qq
\langle J^a(z) J^b(w)\Phi_1\cdots\Phi_N\rangle_{\eta,0}&=&\cr
&&\cr
{_1\over^2}\Biggl(\hskip 2cm
&&
\hskip -2.7cm
\Bigl[\sum_{n=1}^N\omega(z_n,z|\eta) t_{(n)}^a+i\CL_{t^a},
\sum_{n=1}^N\omega(z_n,w|\eta) t_{(n)}^b+i\CL_{t^b}\Bigr]_+
\cr&&
\hskip -3cm
-\sum_{n=1}^N\Bigl(\omega(z_n,z|\eta)[\omega(z,w|\eta) t_{(n)}^b,
t_{(n)}^a]+\omega(z_n,w|\eta)[\omega(w,z|\eta)
t_{(n)}^a, t_{(n)}^b]\Bigr)
\clabel{JaJb}\cr&&
\cr&&
\hskip -3cm
+i\CL_{[t^a,\omega(z,w|\eta)t^b]}+
i\CL_{[t^b,\omega(w,z|\eta)t^a]}
\cr
&&\cr
&&
\hskip -3cm
-k\hspace{.08 cm}\tr\s\Bigl(t^a\partial_z\omega(z,w|\eta)t^b+
t^b\partial_w\omega(w,z|\eta)t^a\Bigr)
\hskip .2cm\Biggr)\hskip .1cm
\langle\Phi_1\cdots\Phi_N\rangle_{\eta,0}\label{2point}
\qqq
where $\omega$ in action on $t^a_{(n)}$ should be
taken in the same representation, $[\ ,\ ]_+$ stands for
the anticommutator and  $\CL_X$ for the Lie
derivative in the direction of the element $X$ of the Lie
algebra of $G$, $\CL_Xf(\eta)=d/dt|_{t=0}f(\ee^{tX}\eta)$.
\vskip 2mm

The energy-momentum tensor of the WZW theory is given by
the Sugawara construction \cite{Suga}:
\qq
(k+h^\vee)T(w)\ =\ {\displaystyle\lim_{x\rightarrow {w}}}\Bigl(
\sum_a J^a(z)J^a(w)
-{_{k\hspace{0.05 cm}{\rm dim}G}\over^{2(z-w)^2}}\Bigr).\label{Sug}
\qqq
Using (\ref{2point}) specialized to the $SU(2)$ case and to
$\eta_u=\exp(2\pi i u\sigma_3)$, we obtain after a straightforward
algebra explained in Appendix B ($t^a\equiv{_1\over^2}\sigma_a,\
t^\pm\equiv t^1\pm it^2$):
\qq
&&
%\displaystyle
\langle T(w)\s\Phi_1\cdots\Phi_N
\rangle_{\eta_u,0}
={_{1}\over^{k+2}}\Bigl[\Bigl(\sum_{n=1}^N
\omega(z_n,w|\eta_u)\s t_{(n)}^3
+ {_{1}\over^{4\pi}}\p_u\Bigr)^2
\cr&&\ +{_{1}\over^{2}}\Bigl[\sum_{n=1}^N
\Bigl(\omega(z_n,w|\eta_u)+{i\over
1-\ee^{4\pi iu}}\Bigr)t_{(n)}^+\ ,\
\sum_{n=1}^N\Bigl(\omega(z_n,w|\eta_u)+{i\over
1-\ee^{-4\pi iu}}\Bigr)t_{(n)}^-\Bigr]_+
\cr
&&\ +i\Bigl(1+2\sum_{r=0}^\infty{q^r\over{\rm e}^{4\pi i u}-q^r}-
2\sum_{r=1}^\infty{q^r\over{\rm e}^{-4\pi i u}-q^r}\Bigr)
\Bigl(\sum_{n=1}^N \omega(z_n,w|\eta_u)\s t_{(n)}^3
+ {_{1}\over^{4\pi}}\p_u\Bigr)
\cr&&\
-k \sum_{r=1}^\infty\Bigl({rq^r\over 1-q^r}
+ {rq^r\over{\rm e}^{4\pi i
u}-q^r}+ {rq^r\over{\rm e}^{-4\pi i u}-q^r}\Bigr)
+{_{3k}\over^{24}}\Bigr]
\s\s\langle\Phi_1\cdots\Phi_N
\rangle_{\eta_u,0}\s.\label{JJ}
\qqq
\vskip 0.4 cm

On the other hand, the Green functions involving the insertion of
the energy-momentum tensor and of the primary fields
follow from the covariance of the theory under diffeomorphisms
$z\mapsto z+\zeta(z)$ such that $\zeta(z+2\pi)=\zeta(z)$
and $\zeta(z+2\pi\tau)=\zeta(z)+2\pi(\tau'-\tau)$. To the linear
order in $\zeta$,
\qq
&\langle\Phi_1(z'_1)\cdots\Phi_N(z'_N)\rangle_{\eta,0}|_{\tau}
+{_1\over^{2\pi i}}\int g_{\bar w\bar w}(w)\hs{0.1 cm}
\langle T(w)\Phi_1(z_1)\cdots\Phi_N(z_N)
\rangle_{\eta,0}|_\tau\hs{0.16 cm}dwd\bar w\cr
&\cr
&=(1+\p_z\zeta(z_1))^{\Delta_1}...(1+\p_z\zeta(z_N))^{\Delta_N}
\s\langle\Phi_1(z_1+\zeta(z_1))\cdots\Phi_N(z_N+\zeta(z_N))
\rangle_{\eta,0}|_{\tau'}\label{cov}
\qqq
where the second term on the left arises because the metric
$dz'd\bar z'$ develops a non-trivial $g_{\bar z\bar z}$ part in new
variables:
\qq
g_{\bar z\bar z}=\p_{\bar z}\zeta.\label{metric}
\qqq
$\Delta_n$ are the conformal weights of fields $\Phi_n$.
Solving eq. (\ref{metric}) for $\zeta$ to the first order in
$g_{\bar z\bar z}$, we obtain
\qq
\zeta(z)={_i\over^{2\pi}}\int\omega(z,w)g_{\bar w\bar w}(w)\hs{0.16
cm}
dwd\bar w,\label{zeta}
\qqq
\vskip -.9 cm
\qq
\tau'=\tau+{_1\over^{4\pi^2}}\int g_{\bar z\bar z}(z)\hs{0.16
cm}dzd\bar z
\label{tau'}
\qqq
where $\omega(z,w)$ is as in (\ref{omega}) but with ${\rm Ad}_\eta$
replaced by $1$. Inserting eqs. (\ref{zeta}),(\ref{tau'})
into (\ref{cov}), we obtain
\qq
&&\langle T(w)\Phi_1\cdots\Phi_N
\rangle_{\eta_u,0}\cr
&&\ \ \ \ \ \ \ =\Bigl(
\sum_{n=1}^N(\Delta_n\p_{w}\omega(z_n,w)
-\omega(z_n,w)\p_{z_n})
-{_1\over^{2\pi
i}}\p_\tau\Bigr)\langle\Phi_1\cdots\Phi_N\rangle_{\eta_u,0}
\label{T}
\qqq
\vskip 0.3 cm

The comparison of the right hand sides of (\ref{JJ})
and (\ref{T}),
gives an identity satisfied by the Green functions.
It may be rewritten as an identity for the holomorphic
factors $\gamma$ we have introduced earlier:
\qq
\nabla(w)\gamma=0,
\label{hor}
\qqq
where
\qq
\nabla(w)&=&
(k+2)\Bigl(\sum_{n=1}^N(-\Delta_n\p_{w}\omega(z_n,w)
+\omega(z_n,w)\p_{z_n})\Bigr)
\cr&+&
\Pi(u,\tau)^{-1}\left\{
{_{k+2}\over^{2\pi i}}\p_\tau+
\Bigl(\sum_{n=1}^N \omega(z_n,w)t_{(n)}^3 +
{_{1}\over^{4\pi}}\p_u\Bigr)^2\right\}\Pi(u,\tau)
\ceqno\cr&+&
{_{1}\over^{2}}\Bigl[\sum_{n=1}^N\Bigl(\omega(z_n,w|\eta_u)+{i\over
1-\ee^{4\pi iu}}\Bigr)t_{(n)}^+\ ,\
\sum_{n=1}^N\Bigl(\omega(z_n,w|\eta_u)+{i\over
1-\ee^{-4\pi iu}}\Bigr)t_{(n)}^-\Bigr]_+
\qqq
and
\qq
\Pi(u,\tau)=q^{1/8}\sin(2\pi u)\prod_r (1-q^r)(1-e^{4\pi iu}q^r)
(1-e^{-4\pi iu}q^r)\label{bigPi}
\qqq
corresponds to
the solution of (\ref{hor}) for the case of $k=2$ and only one
insertion
of spin $j=1$.

Let us discuss the structure of eq. (\ref{hor}). Since $\gamma$
is $w$-independent, the left hand side is, {\it a priori}, a
meromorphic
function of $w$ with poles of at most second order at the insertion
points
$z_n$. Equating the coefficients at $(w-z_n)^{-2}$ to zero, we infer
that
$\Delta_n=j_n(j_n+1)/(k+2)$. Similarly, from the vanishing of the
residues, we obtain equations for the derivatives $\p_{z_n}\gamma$.
They generalize the KZ equations to the toroidal
case. Finally, the regular part of the left-hand side of eq.
(\ref{hor})
is holomorphic and periodic in $w$ so constant and it gives the
derivative $\p_\tau\gamma$.
\vskip 0.3 cm

Explicitly, the different components of the connection are:
\qq\label{connz}
\nabla_{z_n}&=&
\partial_{z_n}
+{_1\over^{2\pi(k+2)}}\Pi(u,\tau)^{-1}
{\partial_u}\Pi(u,\tau)\s t_{(n)}^3\cr&&
+{_1\over^{k+2}}\sum_{m\not=n}
\Big(t_{(n)}^3 (2\s\omega(z_m,z_n)+1)t_{(m)}^3
+t_{(n)}^-( \omega(z_m,z_n|\eta_u)+{i\over 1-\ee^{4\pi iu}})
t_{(m)}^+\cr
&&\hskip 1cm
+t_{(n)}^+( \omega(z_m,z_n|\eta_u)+{i\over 1-\ee^{-4\pi iu}})
t_{(m)}^-\Big)
\qqq
and
\qq\label{contau}
\nabla_\tau&=&\Pi(u,\tau)^{-1}\left(
\partial_\tau+{_i\over^{8\pi(k+2)}}\partial_u^2
\right)\Pi(u,\tau)
\cr&+&
{_{2\pi}\over^{k+2}}\sum_{m,n=1}^N\Big(
t_{(n)}^3 \tilde\omega(z_m,z_n|\eta_u)t_{(m)}^3
+{\textstyle{1\over2}}[t_{(n)}^-\ ,\ (\tilde\omega(z_m,z_n|\eta_u)+
{i\ee^{4\pi iu}\over (1-\ee^{4\pi iu})^2})
t_{(m)}^+]_+\Big)
\qqq
where
\qq
\tilde\omega(z,w|\eta)=
\sum_{r=1}^\infty\left(
{iq^rAd_{\eta^{-1}}\over(Ad_{\eta^{-1}}-q^r)^2}\ee^{ir(w-z)}
+{iq^rAd_\eta\over(Ad_\eta-q^r)^2}\ee^{ir(z-w)}
\right).
\qqq

\noindent To arrive at expressions (\ref{connz}) and (\ref{contau})
we have used some algebraic relations satisfied by $\omega$
that are listed in Appendix C.
\vskip 0.3cm

Eq. (\ref{hor}) has been obtained by manipulating formal functional
integrals. Now, we would like to prove that $\nabla(w)$ defines a
flat
holomorphic connection in the bundle of toroidal Chern-Simons states.
In analogy to the genus zero case where the KZ equations give
holomorphic connection in the trivial bundle of invariant tensors
\cite{Kohno}, one could expect that $\nabla(w)$ defines a holomorphic
connection already in the vector bundle $W_{\un j}$
with the base-space
$\CT_N$ of points $(\un z,\tau)$ and
the fibers composed of maps $\gamma$ satisfying conditions
(\ref{cond}). {\bf This is}, however, {\bf not the case}:
although formally the connection given by $\nabla(w)$ is flat, it
does not preserve the analyticity (in $u$) of maps
$\gamma$ producing, in general, first order poles at
$u\in(\NZ+\tau\NZ)/2$ when acting on $\gamma$.
This is not a problem if $\gamma$ takes values in
$W_{\un j}^{fr}$ (i.e. if
$\gamma(\un z,\tau)$ satisfy additionally conditions
(\ref{reg1})).
We shall see that in this case
$\nabla(w)\gamma$ is analytic as a function of $u$ and, for fixed
$(\un z,\tau)$  and $w\not=z_n$, belongs again to $W_{\un j}^{fr}$.
Consequently, $\nabla(w)$ will define a flat holomorphic
connection on $W_{\un j}^{fr}$ {\bf provided} that we prove
{\bf that the latter is a holomorphic subbundle of} $W_{\un j}$.
This is precisely why we need to define the connection first
on a bigger bundle. Given such a connection with parallel
transport preserving conditions (\ref{reg1}), it would
follow that the spaces of solutions of (\ref{reg1})
have dimension independent of $(\un z,\tau)$ which is
all we need in order to show that $W_{\un j}^{fr}$ is a
subbundle of $W_{\un j}$. It is possible to obtain from
$\nabla(w)$ a well defined connection enlarging
bundle $W_{\un j}$ by admitting $\gamma$'s with poles
at $u=(\NZ+\tau \NZ)/2$. This gives,
however, a connection in the bundle with an
infinite-dimensional fiber where the existence
of the parallel transport does not come for free
(it becomes a hard PDE problem rather than an easy
ODE one). Instead, we shall stay in the finite-dimensional
setup but modify
$\nabla(w)$ to $\tilde\nabla(w)$ by subtracting the pole term
so that the modifications disappear in the action
on $W_{\un j}^{fr}$-valued $\gamma$'s and that
$\tilde\nabla(w)$ gives a well defined (holomorphic)
connection on $W_{\un j}$.

\vs 0.2 cm

First, let us show that if $\gamma$ takes values $W_{\un j}^{fr}$
then $\nabla(w)\gamma$, as a function of $u$, has no poles
at $u=0$. This may be checked directly but may be also
understood by the following argument. If $\gamma$ satisfies
conditions (\ref{reg1}) then it defines a global state $\Psi$. In
particular, for $M\in sl(2,\NC)$ and $A^{01}_M$ as in (\ref{const}),
we may write
$$\Psi(A^{01}_M)=\ee^{{_1\over^2}\pi k\hs{0.05 cm}
{\rm tr}\hs{0.05 cm}(M^2)}\hs{0.1 cm}
\prod_n\left(\ee^{\sigma_3(\bar z_n-z_n)/(2\tau_2)}\right)_{(n)}
\ \gamma(M)$$
so that, with a slight abuse of notation, $\gamma(u)$ coincides
with $\gamma(M)$ for $M=u\sigma_3$, see (\ref{blocks1}).
Due to (\ref{reg1}), $\gamma(M)$ is analytic in $M$.
Besides, it depends only on $\ee^{2\pi iM}\equiv\eta$. Indeed,
if $\ee^{2\pi iM_1}=\ee^{2\pi iM_2}$ then $A^{01}_{M_1}=^h\hs{-0.2cm}
A^{01}_{M_2}$ for $h(z)=\exp[{_1\over^2}\tau_2^{-1}(\bar z-z)M_1]
\exp[{_1\over^2}\tau_2^{-1}(z-\bar z)M_2]$. Application of the
basic covariance property (\ref{blocks}) of the states gives the
equality
$\gamma(M_1)=\gamma(M_2)$. Since the exponential map
$M\mapsto\eta$ may be locally inverted around $M=0$,
it follows that $\gamma(\eta)$ is analytic around $\eta=1$
(in fact it is even globally analytic on $SL(2,\NC)$).
Now the terms in $\nabla(w)$ with the poles at $u=0$ came from
rewriting $\CL_{t^\pm}$ and $\CL_{t^\pm}\CL_{t^\mp}$ in (\ref{JaJb}),
see Appendix C. In $\nabla(w)\gamma(u)$, they may be reabsorbed
into $\CL_{t^\pm}\gamma(\eta)|_{\eta=\eta_u}$,
$\CL_{t^\pm}\omega(z_n,w|\eta_u)$ and
$\CL_{t^\pm}\CL_{t^\mp}\gamma(\eta)|_{\eta=\eta_u}$, all
three regular in $u$ around $u=0$.
\vskip 2mm

To show that the other poles in $u$ are absent in $\nabla(w)\gamma$
when $\gamma$ satisfies conditions (\ref{reg1}), we shall
use operators $\CQ_i$, $i=1,2$, essentially extending the action
(\ref{blocks}) of gauge transformations to the  case of ``almost''
gauge transformations $h_v$ for $v={_1\over^2}$
and $v={_\tau\over^2}$, see (\ref{h}) (for the later convenience
we shall use the subindex ${\un j}$, in the notation
for $\CQ_i$'s, to keep track of the insertions).
\qq
\CQ_{{\un j},1}\s\gamma(u)&=&\gamma(u+1/2)\s,
\cr
\CQ_{{\un j},2}\s\gamma(u)&=& \ee^{2\pi i k(u+\tau/4)}
\prod_n\ee^{-iz_nt^{3}_{(n)}}\gamma(u+\tau/2)
\qqq
and they act at the space of analytic
maps $\NC\setminus(\NZ+\tau\NZ)/2\ni u\mapsto
\gamma(u)\in \otimes V_{j_n}$. We shall also need
\qq
\hbox to 1.5cm{$\CP_{{\un j},i}$\hfill}&=&\CQ_{{\un
j},i}^2\qquad\qquad
{\rm for}\ i=1,2\s,
\cr
\hbox to 1.5cm{$\CP_{{\un j},3}$\hfill}&=&\prod_n \ee^{it_{(n)}^3}\s,
\cr
\hbox to 1.5cm{$\CP_{{\un j},4}\gamma(u)$\hfill}&=&\prod_n
\Omega_{(n)}
\gamma(-u)\s,
\qqq
with $\Omega$ the generator of the Weyl group of $SL(2,\NC)$.

These operators satisfy the algebra
\qq
\hbox to 1.9cm{$[\CQ_{{\un j},1},\CQ_{{\un j},2}]$\hfill}&=&0\s,\cr
\hbox to 1.9cm{$[\CQ_{{\un j},i},\CP_{{\un j},3}]$\hfill}
&=&0\s,\clabel{comm}\cr
\hbox to 1.9cm{$\CQ_{{\un j},i}\CP_{{\un j},4}$\hfill}
&=&\CP_{{\un j},4}\CQ_{{\un j},i}^{-1}\s,\cr
\hbox to 1.9cm{$\CP_{{\un j},3}\CP_{{\un j},4}$\hfill}
&=&\CP_{{\un j},4}\CP_{{\un j},3}^{-1}
\qqq
and conditions (\ref{cond}) can be written as
\qq
(\CP_{{\un j},i}-1)\gamma = 0\hbox{\qquad for }\
i=1,\dots,4\s.\label{condn}
\qqq
Defining
\qq
D^L_{{\un j},\alpha}=\hat D^L_{{\un j},0}\CQ_{{\un j},1}^a\CQ_{{\un
j},2}^b\qquad
\hbox{for}\ \
\alpha\equiv(a+b\tau)/2=0,1/2,\tau/2,(\tau+1)/2\s,\label{regn}
\qqq
we may rewrite the conditions
of (\ref{reg1}) in an equivalent way using operators
$D^L_{{\un j},\alpha}$ instead of
$\hat D^L_{{\un j},\alpha}$.
{}From (\ref{comm}), (\ref{condn})  and (\ref{regn})
it is clear that $\CQ_{{\un j},i}$ preserve spaces $W_{\un j}(\un
z,\tau)$
and $W_{\un j}^{fr}(\un z,\tau)$. Also
\qq
&&[\CQ_{{\un j},i},\nabla(w)]=0\ \ \ \ {\rm for}\ i=1,2\s,
\cr &&
[\CP_{{\un j},i},\nabla(w)]=0\ \ \ \
{\rm for}\ i=1,\cdots,4\s.
\qqq
so that
\qq
\nabla(w)\s\gamma(u+(p+r\tau)/2)&=&
\ee^{-2\pi ik(ru+r^2\tau/4)}\otimes_n\ee^{inz_nt_{(n)}^3}\CQ_{{\un
j},1}^p\CQ_{{\un j},2}^r
\nabla(w)\gamma(u) \cr
&=&\ee^{-2\pi ik(ru+r^2\tau/4)}\otimes_n\ee^{inz_nt_{(n)}^3}
\nabla(w)\CQ_{{\un j},1}^p\CQ_{{\un j},2}^r\gamma(u).
\qqq
As $\nabla(w)\gamma$ is analytic around $u=0$
provided $\gamma$ is a Chern-Simons state, it follows that,
in this case, $\nabla(w)\gamma$ is entire as a function of $u$.

Let us pass to redefining $\nabla(w)$ by
subtracting ``by hand'' its poles in such a way
that the modifications vanish when acting on $W_{\un j}^{fr}$-valued
$\gamma$'s and the new regular $\tilde\nabla(w)$ defines a
holomorphic connection in $W_{\un j}$. This may be done by means of
operators $R_n$, $R_0$,
\qq
R_n\gamma (u)&=&
{i\over4}\sum_{r\in\NZ}\ee^{2\pi i k(r^2\tau-2ru)}
\left({q^r\over\ee^{2\pi iu}-q^r}
-{q^{-r}\over\ee^{-2\pi iu}-q^{-r}}\right)
\cr &&
\qquad\qquad(\prod_m\ee^{2irz_mt^3_{(m)}})
(t_{(n)}^+{\textstyle \sum_m t_{(m)}^-}-t_{(n)}^-
{\textstyle \sum_m t_{(m)}^+})
\s\gamma(0)\qquad\hbox{for}\quad n=1,\dots,N\s,
\cr\cr
R_0\gamma (u)&=&
{1\over8}\sum_{r\in\NZ}{\ee^{2\pi i k(r^2\tau-2ru)}
\over(\ee^{2\pi iu}-q^r)(\ee^{-2\pi iu}-q^{-r})}
({\prod_m}\ee^{2irz_mt^3_{(m)}})
[{\textstyle \sum_m t_{(m)}^+},{\textstyle \sum_m t_{(m)}^-}]_+
{\gamma(0)}\clabel{rreg}\cr
\cr&+&{i\over32\pi}\sum_{r\in\NZ}\ee^{2\pi i k(r^2\tau-2ru)}
\left({q^r\over\ee^{2\pi iu}-q^r}
-{q^{-r}\over\ee^{-2\pi iu}-q^{-r}}\right)
\cr
\cr&&\qquad\qquad({\prod_m}\ee^{2irz_mt^3_{(m)}})
([{\textstyle \sum_m t_{(m)}^+},{\textstyle \sum_m t_{(m)}^-}]_+-4)
{\p_u\gamma(0)}
\qqq
which vanish on $W_{\un j}^{fr}(\un z,\tau)$ (see Appendix D) and
make
\qq
\tilde\nabla(w)=\nabla(w)-
\sum_{a,b=0,1}\CQ_{{\un j},1}^a\CQ_{{\un j},2}^b
\left(
{_1\over^{2\pi i}} R_0
-\sum_{n=1}^N \omega(z_n,w) R_n
\right)\CQ_{{\un j},1}^{-a}\CQ_{{\un j},2}^{-b}
\qqq
regular when acting on $W_{\un j}$.
Note that $(\CP_i -1)R_n =0$ for $i=1,\dots,4$ and $n=0,\dots,N$
so that $\tilde\nabla(w)$ actually defines a connection on
$W_{\un j}$. This connection is not flat any more but, as may be
checked, its curvature vanishes on $W_{\un j}^{fr}$-valued sections.

Now we want to show that the connection on $W_{\un j}$ defines by
parallel
transport an isomorphism between spaces  $W^{fr}_{\un j}(\un z,\tau)$
for different
$(\un z,\tau)$. This comes from the following relations obtained
by commuting $D^L_{{\un j},\alpha}$ and $\tilde\nabla(w)$
or its components $\tilde\nabla_{z_n}$, $\tilde\nabla_\tau$
(a long but straightforward calculation which is detailed
in Appendix D)
\qq
D^L_{{\un j},\alpha}\tilde\nabla_{z_n}\gamma&=&
\p_{z_n}D^L_{{\un j},\alpha}\gamma+\sum_{\beta,L'}
A_{n,}{}_{\alpha L'}^{\beta L}D^{L'}_{{\un j},\beta}\s\gamma
\ \ \ \ \ {\rm for} &\ $n=1,\dots,N$\s,\hskip 1.7cm\cr&\clabel{DLC}\cr
D^L_{{\un j},\alpha}\tilde\nabla_\tau\gamma&=&
\p_{\tau}D^L_{{\un j},\alpha}\gamma+\sum_{\beta,L'}
A_{0,}{}_{\alpha L'}^{\beta L}D^{L'}_{{\un j},\beta}\s\gamma\s,
\qqq
where for $|L|<J$ and $n=0,\dots,N$,  $A_{n,}{}_{\alpha L'}^{\beta
L}$
is different from zero
only if $|L'|<J$.
Thus if $D^L_{{\un j},\alpha}\gamma$ vanish for $|L|<J$ at the
initial
point of the $\tilde\nabla$-horizontal curve in $W_{\un j}$ they
also do along the whole curve. As a consequence,
dimensions of spaces $W_{\un j}^{fr}(\un z,\tau)$ are independent of
points $(\un z,\tau)$ and
$W_{\un j}^{fr}$ is a holomorphic subbundle
of $W_{\un j}$ with a holomorphic (flat) connection $\nabla$
coinciding with the restriction of
$\tilde\nabla$.

Using the properties of $\omega(z,w|\eta)$ under modular
transformations, listed in Appendix C, it is easy
to check that the action of the modular group $\Gamma_N$
almost preserves the connection $\tilde\nabla$:
$\tilde\nabla_{z_n}$ is preserved but the pullback of
$\tilde\nabla_\tau$ is shifted by $c(c\tau+d)^{-1}\sum_n\Delta_n$
(not very precisely, we may say that $\gamma$'s transform as
the product  of $\Delta_n$-forms). As the result,
$\tilde\nabla$ and $\nabla$ define canonically only projective
connections on $W_{\un j}/\Gamma_N$ and
$W_{\un j}^{fr}/\Gamma_N$, respectively.
\vskip .5cm
\nsection{\hspace{-.7cm}.\ \ Factorization}

In this Section we would like to derive relations between
the spaces of blocks with different number of insertions.
In order to do so we shall study, as in \cite{GK}, the behavior of
our bundle when two punctures come together. Taking
the punctures at $(z_1=z_2+\zeta^{2(k+2)},z_2,\cdots,z_N)$,
we shall let $\zeta\to 0$. We shall study the case
when $j_1=1/2$ and the rest are positive integrable
spins i.e. $0<j_n\leq k/2\ n=2,\cdots,N$.

Let us consider the connection
\qq
\nabla_{\zeta}=
{\zeta^{2k+3}\over\pi i}\oint_{C_1} {dw\over w} \tilde\nabla(w)
\qqq
where the path $C_1$ encloses clockwise only insertion at
$z_1$\s.
\qq
\nabla_{\zeta}= \partial_\zeta-2\zeta^{-1}(2 t^3_{(1)}t^3_{(2)}
+t^+_{(1)}t^-_{(2)}+t^-_{(1)}t^+_{(2)})+\CO(\zeta^{2k+3})
\qqq
and the connection has a singularity when $\s\zeta\to 0\s$.
\s To regularize it we must diagonalize the
singular term. The decomposition
\qq
\bigotimes_n V_{j_n}\cong\bigoplus_{j=j_2\pm1/2}
{Inv}(V_{1/2}\otimes V_{j_2}\otimes V_j)\otimes
(V_j\otimes V_{j_3}\cdots\otimes V_{j_N})
\qqq
does the job. Explicitly,
\qq
\gamma(u)((v_n)) =
\sum_{j=j_2\pm1/2} \s\sum_{l=0}^{2j}\s\p_v^{2j-l}|_{v=0}\s
P_{CG}^{1/2j_2j}(v_1,v_2,v)\s\partial_{v'}^l|_{v'=0}\s
\hat\gamma_{jj_3\cdots j_N}(u) (v',v_3,\cdots,v_N)
\qqq
where the Clebsch-Gordan invariant tensors are given by
\qq
P_{CG}^{j_1j_2j_3}(v_1,v_2,v_3)
=(v_1-v_2)^{j_{12}}(v_1-v_3)^{j_{13}}(v_2-v_3)^{j_{23}}
\qqq
if $j_{12}=j_1+j_2-j_3$,
$j_{13}=j_1+j_3-j_2$ and
$j_{23}=j_2+j_3-j_1$ are non-negative integers and by zero otherwise.

Now we perform a gauge transformation in the $\hat\gamma$'s
by the isomorphism (at $\zeta\neq 0$)
\qq
\varphi_{_\zeta}\gamma &=&
(\tilde\gamma_{{\un j}^+}\ ,\
\tilde\gamma_{{\un j}^-})
\cr&=&
(\zeta^{-2j_2}\hat\gamma_{{\un j}^+}\ ,\
\zeta^{2j_2+2}\hat\gamma_{{\un j}^-}).
\label{gauge1}
\qqq
where ${\un j}^{\pm}$ stands for insertions of spins
$j_2\pm 1/2,j_3,\dots,j_N$.
In this way the connection $\tilde\nabla_\zeta=
\varphi_{_\zeta}^{-1}\nabla_\zeta\varphi_{_\zeta}$
becomes regular at $\zeta=0$.

On the space of states we have
\qq
\varphi_{_\zeta}  W_{\un j}^{fr}
((z_2+\zeta^{2(k+2)},z_2,\cdots,z_N),\tau)
=
\tilde W^{fr}_{{\un j}}(\zeta)\s.
\qqq
By the parallel transport with
$\tilde\nabla_\zeta$
to $\zeta=0$, we obtain  space $\tilde W_{{\un j}}^{fr}(0)$.
Our next goal will be to show that this space coincides
with $\s W_{{\un j}^+}^{fr}((z_2,\cdots,z_N),\tau)
\oplus W_{{\un j}^-}^{fr}((z_2,\cdots,z_N),\tau)\s$.
Let $\s(\tilde\gamma_{{\un j}^+}\s,\s\tilde\gamma_{{\un j}^-})\s$,
defined for small $|\zeta|$,
be $\tilde\nabla$-horizontal and with values
in $\s\tilde W_{{\un j}}^{fr}(\zeta)\s$.
\vskip 0.2cm

First $\s(\CP_{{\un j}^{\pm},i}-1)\s
\tilde\gamma_{{\un j}^\pm}(\zeta)=0\s$
for $i=1,3,4$ and all $\zeta\not=0$ so,
by continuity, also for $\zeta=0$.
As for $\CP_{{\un j},2}$ we have:
\qq
\CP_{{\un j},2}&=&  \ee^{2\pi i k(2u+\tau)}\ee^{{\tau}\p_u}
\exp[-2i(z_2+\zeta^{2(k+2)}){t_{(1)}^3}-2iz_2{t_{(2)}^3}-
\cdots-2iz_N{t_{(N)}^3}]
\cr&=&
\tilde\CP_{{\un j},2} + \CO(\zeta^{2(k+2)})
\qqq
where the $\CO$-term is a linear operator and
\qq
\tilde\CP_{{\un j},2}=\ee^{2\pi i k(2u+\tau)}\ee^{{\tau}\p_u}
\exp[-2i z_2{(t_{(1)}^3+ t_{(2)}^3)}-2i z_3{t_{(3)}^3}
-\cdots-2i z_N{t_{(N)}^3}]\s.
\qqq
Then
\qq
&&\varphi_{_\zeta}  (\tilde\CP_{{\un j},2}-1)\s
\gamma((z_2+\zeta^{2(k+2)}, z_2,\cdots, z_N),\tau)
\cr
&=&
(\s(\CP_{{\un j}^+,2}-1)\s\tilde\gamma_{{\un j}^+}(\zeta)\s,\s
(\CP_{{\un j}^-,2}-1)\s\tilde\gamma_{{\un j}^-}
(\zeta)\s)=\CO(\zeta^{2(k+1-
j_2)})
\qqq
and $\s(\CP_{{\un j}^\pm,2}-1)\s\tilde\gamma_{{\un j}^\pm}(0)=0\s$.

We shall study the rest of the conditions that define the
space of states
i.e. those involving operators $D_{{\un j},\alpha}^L$.
Due to the homogeneity of maps $\gamma$, we may always take
$L$ in $D_{{\un j},\alpha}^L$ with one of the $l_n=0$
and obtain an equivalent
set of conditions. In the following we will take $L$ with $l_2=0$ and
to remember this restriction we shall label such $L$ as $\tilde L$.

Now we are going to examine the behavior of
$D_{{\un j},\alpha}^{\tilde L}\gamma$ near
$\zeta=0$. Let $L^n$ ($L_n$) denote multi-index $L$ with $l_n$
increased (lowered) by $1$ (if $l_n=0$ then terms with $L_n$ should
be omitted).
\qq
D^{\tilde L}_{{\un j},\alpha}\nabla_\zeta\gamma=
\partial_\zeta D_{{\un j},\alpha}^{\tilde L}\gamma - 2\zeta^{-1}
\Bigl[&&\hskip -3mm
(j_2(1-2l_1)-l_1(2-l_1))
D_{{\un j},\alpha}^{\tilde L}\gamma
\clabel{connD}\cr
- &&\hskip -3mm
l_1(2-l_1) \Bigl((j_2+1/2)D_{{\un j},\alpha}^{\tilde L_1}\gamma-
\sum_{n=3}^N ( D_{{\un j},\alpha}^{\tilde L_1^n}\gamma
+ (j_n-l_n) D_{{\un j},\alpha}^{\tilde L_1}\gamma)\Bigr)
\Bigr]
\cr
+\CO(\zeta^{2k+3})\hskip -4mm
\qqq
which develops a singularity when $\zeta\to 0$.
We have used the homogeneity of degree $J$ in variables
$(v_n)$ of polynomial $\gamma$ and so:
$$\sum_n ( D_{{\un j},\alpha}^{L^n} + (l_n - j_n) D_{{\un
j},\alpha}^L)\s\gamma = 0\s. $$
The transformation
\qq
\tilde D_{{\un j},\alpha}^{\tilde L}\gamma&=& \zeta^{-2j_2} D_{{\un
j},\alpha}^{\tilde L}\gamma
\hskip 8.6cm{\rm for}\ \ l_1=0\s,
\cr
\cr
\tilde D_{{\un j},\alpha}^{\tilde L}\gamma&=& \zeta^{2(j_2 +1)}
\Bigl(D_{{\un j},\alpha}^{\tilde L}\gamma + {_1\over^2} D_{{\un
j},\alpha}^{\tilde L_1}\gamma
\ceqno\cr
&&
\hskip 1.45cm
- (2j_1+1)^{-1}
\sum_{n=3}^N  (D_{{\un j},\alpha}^{\tilde L_1^n}\gamma
+ (j_n-l_n) D_{{\un j},\alpha}^{\tilde L_1}\gamma)\Bigr) \hskip
1.4cm{\rm for}\ \ \s l_1=1
\qqq
removes the singularity in (\ref{connD}).
Now if $\gamma$ is horizontal with respect
to $\nabla_\zeta$ around $\zeta=0$, we have
\qq
{d\over d\zeta}\tilde D_{{\un j},\alpha}^{\tilde L}\gamma =
\sum_{\beta,\tilde L'}
B_{\alpha\tilde L'}^{\beta\tilde L}
\tilde D_{{\un j},\beta}^{\tilde L'} \gamma
\label{horD}
\qqq
with $B_{\alpha L'}^{\beta\tilde L}$ analytic at $\zeta=0$. Consequently,
we can extend $\tilde D_{{\un j},\alpha}^{\tilde L}\gamma(\zeta)$ to
$\zeta=0$.

Let us introduce
\qq
\tilde\CQ_{{\un j},2}&=&
\ee^{2\pi i k(u+\tau/4)}\ee^{{\tau\over 2}\p_u}
 \exp[-iz_2{(t_{(1)}^3+t_{(2)}^3)}-i z_3{t_{(3)}^3}
-\cdots-i z_N{t_{(N)}^3}]
\cr
&=&
\CQ_{{\un j},2}+\CO(\zeta^{2(k+2)})
\qqq
with $\s\varphi_{_\zeta}\tilde\CQ_{{\un j},2}=
(\CQ_{{\un j}^+,2}\oplus\CQ_{{\un j}^-,2})\s\varphi_{_\zeta}\s$.
\s In terms of states $\tilde\gamma_{{\un j}^\pm}$ of (\ref{gauge1})
and for $\alpha=(a_1+\tau a_2)/2,\  a_i=0,1$,
operators $\tilde D_{{\un j},\alpha}^{\tilde L}$
are
\qq
\tilde D_{{\un j},\alpha}^{\tilde L}
\gamma &=&\tilde D_{{\un j},0}^{\tilde L}
\CQ_{{\un j},1}^{a_1}
\tilde\CQ_{{\un j},2}^{a_2}\gamma+\CO(\zeta^{2(k+1-2j_2)})\cr
&=&(2j_2+1)! D_{{\un j}^+,\alpha}^{\tilde L} \tilde\gamma_{{\un j}^+}
\cr&&
+(2j_1)!\zeta^{2(k+2)} D_{{\un j}^+,\alpha}^{\tilde L}\partial_{v'}
\tilde\gamma_{{\un j}^+}
\ceqno\cr&&
+(2j_1-1)\zeta^{2(k+1-2j_2)} D_{{\un j}^-,\alpha}^{\tilde L}
\tilde\gamma_{{\un j}^-}
\cr&&+\CO(\zeta^{2(k+1-2j_2)})&\hspace{-0.5cm}for $l_1=0$\hskip 1cm\cr
\qqq
\vskip -\baselineskip
\no and
\vskip -\baselineskip
\qq
\tilde D_{{\un j},\alpha}^{\tilde L}\gamma&=&
\tilde D_{{\un j},0}^{\tilde L}
\CQ_{{\un j},1}^{a_1}
\tilde\CQ_{{\un j},2}^{a_2}\gamma+\CO(\zeta^{2(k+2)})
\cr&=&(2j_2-1)! D_{{\un j}^-,\alpha}^{\tilde L_1} \tilde\gamma_{{\un
j}^-}
+(2j_1)!\zeta^{2(2j_2+1)} D_{{\un j}^+,\alpha}^{\tilde
L_1}\partial_{v'}
\tilde\gamma_{{\un j}^+}
\ceqno\cr&&
+(2j_1+1)^{-1} \zeta^{2(2j_2+1)}
\sum_{n=3}^N (\tilde D_{{\un j},\alpha}^{\tilde L_1^n}\gamma -
 (j_n-l_n) \tilde D_{{\un j},\alpha}^{\tilde L_1}\gamma)
\cr&&+\CO(\zeta^{2(k+2)})&\hspace{-0.5cm}for $l_1=1\s.$\hskip 1cm\cr
\qqq
But equation (\ref{horD}) implies that $\tilde D_{{\un
j},\alpha}^{\tilde L}\gamma=0$
for any $|\tilde L|<J$ and at any value of $\zeta$. For $\zeta=0$
we obtain:
\qq
D_{{\un j}^+,\alpha}^L\tilde\gamma_{{\un j}^+}=0\ \ \ {\rm for}\ \
&&|L|<j_2
+1/2+j_3+\cdots+j_N\s,\cr&&\alpha=0,1/2,\tau/2,(\tau+1)/2
\qqq
and
\qq
D_{{\un j}^-,\alpha}^L\tilde\gamma_{{\un j}^-}=0\ \ \ {\rm for}\ \
&&|L|<
j_2-1/2+j_3+\cdots+j_N\s,\cr&&\alpha=0,1/2,\tau/2,(\tau+1)/2
\qqq
which are precisely conditions for bundles
$W^{fr}_{{\un j}^\pm}$. This proves that
$$\tilde W_{\un j}^{fr}(0)=\s W_{{\un j}^+}^{fr}((z_2,\cdots,z_N),\tau)
\oplus W_{{\un j}^-}^{fr}((z_2,\cdots,z_N),\tau)\s.$$
\vskip 0.1cm

Denoting $\s\CD_{j_1\cdots j_N}={\rm dim}\s W_{\un j}^{fr}\s$ we thus
obtain
\qq
\CD_{1/2\hspace{0.03cm}j_2j_3\cdots j_N}=\CD_{(j_2+1/2)j_3\cdots j_N}+
\CD_{(j_2-1/2)j_3\cdots j_N}\s.\label{dim1}
\qqq
The formula of \cite{Verl}\cite{MS} that we want to prove is
\qq
\CD_{j_1\cdots j_N}=\sum_{j=0}^{k/2} S_{j_1j}\cdots
S_{j_Nj}/(S_{0j})^N,
\label{Ver}
\qqq
with
$$S_{jj'}=(2/(k+2))^{1/2} \sin[\pi(2j+1)(2j'+1)/(k+2)]\s.$$
Eq. (\ref{Ver}) is a consequence of (\ref{dim}) and of the formal
factorization property
\qq
\CD_{j_1\cdots j_N}=\sum_{j=0}^{k/2}N_{j_1\cdots j_mj}\s
\CD_{jj_{m+1}\cdots j_N},\label{fact}
\qqq
where $N_{j_1\cdots j_N}$ is the dimension of the space of
Chern-Simons states in the spherical topology, see \cite{GK}.

Now, having under control the factorization when one of the
spins is $1/2$, we can, by an inductive procedure,
prove (\ref{Ver}) rigorously, modulo the absence of toroidal states
with spins $>k/2$.
First,
\qq
N_{1/2\hspace{0.03cm}j_2j}=\cases{
\matrix{1 \hfill&&\hbox{for \ }\vert j_2-j\vert=1/2
\hbox{\ and \ }j_2,j\leq k/2\s,\hfill\cr
0\hfill&&\hbox{otherwise\s.}\hfill}}
\qqq
Formula (\ref{dim1}) is a particular case of (\ref{fact})
and applying it iteratively we can show
(\ref{Ver}) for $j_2=\cdots=j_N=1/2$.
For the other cases, we may use an inductive procedure, in the
increasing order of the sums $J$ of inserted spins and, for
equal $J$'s, in the decreasing order of the numbers of insertions.
Take $j_1>1/2$. Then, from (\ref{dim1}),
$$
\CD_{j_1j_2\cdots j_N}=\CD_{1/2 (j_1-1/2) j_2\cdots j_N}-
\CD_{(j_1-1)j_2\cdots j_N},
$$
and on the right-hand side only earlier sequences appear
so that we may proceed inductively. This proves formula
(\ref{Ver}) for the dimension of spaces $W_{\un j}^{fr}$ of $SU(2)$
Chern-Simons states on the torus.

\nsection{\hspace{-.7cm}.\ \ Conclusions}

Let us summarize the main results of the paper. First
we have built a stratification of the space of
rank two, topologically trivial, holomorphic bundles on the torus.
This has been a preliminary
step in the description of the states of the $SU(2)$ CS theory
in this geometry
as polynomials with theta-functions of degree  $2k$ as coefficients.
We have constructed the toroidal Knizhnik-Zamolodchikov
connection and have shown that the parallel transport preserves
the spaces of states. Modulo the integrability
condition (if one of spins $>k/2$ then there are no non-trivial
states), we have proven the factorization property
of the spaces of states which implies the Verlinde formulae for
the dimension of the space of toroidal conformal blocks.

As was remarked before, the integrability condition
should follow from the definition of the space of states
as it does in the spherical topology or for the toroidal
one-point blocks. We expect to derive it for toroidal
$N$-point blocks by extending the bundles of states to the
point where the torus degenerates into the sphere
(i.e. when $\tau$ goes to $i\infty$).
We will address this problem in the future.

It is also interesting to understand the scalar product turning
the space of Chern-Simons states into a Hilbert space.
In the context of the conformal field theory,
the scalar product of states gives the
pairing of holomorphic and antiholomorphic conformal blocks into
correlation functions. It should be determined uniquely up
to normalization by demanding its invariance under the parallel
transport with respect to the Knizhnik-Zamolodchikov connection.
Up to now we know expressions for the scalar product at genus
zero and one in terms of
multiple integrals (see \cite{G}\cite{FalGK} for the spherical case).
We expect the integrals to
converge if (and only if) we apply them to Chern-Simons
states. The proof of this conjecture (still absent for the
spherical topology, too) should be the goal of the future research.

\vskip 1cm

\nappendix{A}
\vskip 0.5cm

\noindent In this Appendix we complete the proof
of the stratification of Section 3. We refer the reader
to that Section for the notations used here.
We shall show that the map of (\ref{map})
\qq
P_0:\CG_0^\NC\times U_0&\longrightarrow& \CA^{01}\cr
(g,A_M^{01})&\longmapsto& {}^g A_M^{01}
\qqq
is smooth and injective, its differential is
invertible everywhere and its inverse is $C^\infty$ when considered
as a map
$$(\CG_0^\NC\times U_0)\times \CA^{01}\longrightarrow
T(\CG_0^\NC\times U_0).$$
Then, by the Nash-Moser Theorem \cite{Ham},
$P_0$ is a $C^\infty$ diffeomorphism in the Fr\'echet sense
from $\CG_0^\NC\times U_0$ onto its open image.
\vskip 0.1cm

It is clear that $P_0$ is
smooth and tame in the $C^\infty$ Fr\'echet structure of
$\CG_0^\NC\times U_0$ and
$\CA^{01}$.
To prove that $P_0$ is injective one has to consider the equation
\qq
{}^h{\hskip -1mm} A^{01}_{M'}=A^{01}_M&&{\rm for}\
h\in\CG_0^\NC\ \ {\rm and }\ \
A^{01}_M,\ A^{01}_{M'}\in U_0
\label{inj}
\qqq
where $A^{01}_M,\ A^{01}_{M'}$ are like in (\ref{const}).
The general solution of (\ref{inj}) is
\qq
h(z,\bar z)=\exp(i{z-\bar z\over\tau-\bar\tau}M)\s g(z)
\exp(-i{z-\bar z\over\tau-\bar\tau}M')
\qqq
with
$g:\NC\rightarrow SL(2,\NC)$ holomorphic in the complex plane and
\qq
g(z+2\pi)&=&g(z)\s,\alabel{per}a\cr
g(z+2\pi\tau)&=&\exp(-2\pi i M) g(z) \exp(2\pi i M')\s.\aeqno b\cr
\qqq

Upon expanding $g$ in Fourier modes,
$$g(z)=\sum_{r\in\NZ}g_r\exp({2\pi i r z})\s,$$
eq. (\ref{per},b) becomes equivalent to the statement
that the $2\times 2$-matrices
$g_r$ are eigenvectors of
$$g_r\ \smash{\mathop{\mapsto}\limits_{}^{\chi}}
\ \displaystyle \exp({-2\pi i M})\s\s g_r
\s\exp({2\pi i M'})$$
with eigenvalues $\exp({2\pi i r\tau})$.
But eigenvalues of $\chi$ are $\exp[\pm 2\pi i(\det^{1/2}M\pm
\det^{1/2}M')]$ and, provided (\ref{bound}) is satisfied,
the only non trivial solution comes from $\det M=\det M'$ and $r=0$.
This, together with $h(0)=1$, shows that the only solution of
(\ref{inj})
is $h=1$ and, consequently, that $P_0$ is injective.
\vskip 0.1cm

Now we prove that the differential $DP_0$ is invertible and its
inverse
is a smooth tame map.
First note that, with the use of left invariant vector fields
to describe vectors in $T\CG_0^\NC$,
\qq
DP_0((g,A_M^{01}))(\phi,B)=g\left(-\bar\partial\phi
-[A_M^{01},\phi]+
B\right)g^{-1}\equiv g \psi g^{-1}\s.
\qqq
Using the expansion in modes
$$\psi(z,\bar z)=\sum_{p,r\in\NZ}
\psi_{p,r}\exp[i(r(z-\bar z)+p(\bar\tau z - \tau \bar z))/(\tau-
\bar\tau)]$$
and the same for $\phi(z,\bar z)$, one obtains
\qq
\psi_{p,r}&=&i{r+p\tau\over\tau-\bar\tau}\phi_{p,r}d\bar
z-[A^{01}_M,\phi_{p,r}]
\hskip 1.5cm& for $\ (p,r)\not=(0,0)\s,$\hskip 1cm\alabel{diff} a\cr
\phi_{0,0}&=&-\sum_{(p,r)\not=(0,0)}\phi_{p,r}\s,\aeqno b\cr
B&=&\psi_{0,0}+[A_M^{01},\phi_{0,0}]\s.\aeqno c\cr
\qqq
It is evident that this equations can be solved in $\phi$ and $B$ if
(\ref{diff},a) can, and this is so if $A_M^{01}\in U_0$. One
easily sees, in this case, that the inverse is a smooth tame family
of linear maps, as stated before. We conclude that $P_0$ is
a $C^\infty$ diffeomorphism of Fr\'echet manifolds.
\vskip 1cm

\nappendix{B}
\vskip 0.6 cm

\noindent Let us explain the passage from
eq. (\ref{JaJb}) to (\ref{JJ}).
\vs 0.2 cm
\no The last term inside $\Big(...\Big)$ in (\ref{JaJb}) ,
with the $\hs{0.1 cm}\hs{0.1 cm}\sim(z-w)^{-2}$
singularity subtracted, gives the last two terms
proportional to $k/(k+2)$ in $\Big[...\Big]$ on
the right hand side of (\ref{JJ}).
\vs 0.2 cm

\no The second and the third term in $\Big(...\Big)$ contribute
to the third term in $\Big[...\Big]$ the part with
$\sum_{r=1}^\infty$.
\vs 0.2 cm

The rest of (\ref{JJ}) comes from the first term in $\Big(...\Big)$
of (\ref{JaJb}). The main input is the covariance of Green functions:
$$\langle\Phi_1...\Phi_N\rangle_{{\rm Ad}_\rho\eta_u,0}
=\otimes_nD_{R_n}(\rho)\hs{0.15 cm}\langle\Phi_1...\Phi_N
\hs{0.1 cm}\rangle_{\eta_u,0}
\otimes_nD_{R_n}(\rho^*).$$
Then, since \
$\ee^{\epsilon t^\pm}\eta_u={\rm Ad}_{\rho(\epsilon)}
\eta_u
\hs{0.15 cm}+\hs{0.1 cm}\CO(\epsilon^2)$ where $\rho(\epsilon)=
\epsilon(1-\exp[\pm4\pi iu])^{-1}t^\pm$,
$$\CL_{t^\pm}\langle\Phi_1...\Phi_N\rangle_{\eta_u,0}=
\sum_n(1-\exp[\pm4\pi iu])^{-1}t^\pm_{(n)}
\ \langle\Phi_1...\Phi_N\rangle_{\eta_u,0}\hs{0.1 cm}.$$
Similarly one shows that
$${_1\over^2}\CL_{t^\pm}\CL_{t^\mp}
=\mp(1-\exp[\pm4\pi iu])^{-1}\CL_{t^3}+{_1\over^2}\Big(\sum_n
(1-\exp[\pm4\pi iu])^{-1}t^\pm_{(n)}\Big)\Big(\sum_n
(1-\exp[\mp 4\pi iu])^{-1}t^\mp_{(n)}\Big)$$
when acting on $\langle\Phi_1...\Phi_N\rangle_{\eta_u,0}$.
Finally,
\qq
&&{_1\over^2}\CL_{t^\pm}\hs{0.1 cm}\omega(z_n,w|\eta_u)t^\mp_{(n)}\cr
&&={_1\over^2}
[{\rm ad}_{(1-\exp[\pm 4\pi iu])^{-1}t^\pm},
\omega(z_n,w|\eta_u)]\hs{0.1 cm}
t^\mp_{(n)}+{_1\over^2}\omega(z_n,w|\eta_u)\hs{0.1 cm}t^{\mp}_{(n)}
\sum_{m}(1-\exp[\pm4\pi iu])^{-1}t^\pm_{(m)}\cr
&&={_1\over^2}\sum_{m}(1-\exp[\pm4\pi iu])^{-1}t^\pm_{(m)}
\hs{0.1 cm}\omega(z_n,w|\eta_u)
\hs{0.1 cm}t^{\mp}_{(n)}-{_1\over^2}\omega(z_n,w|
\eta_u)\hs{0.16 cm}{\rm ad}_{(1-\exp[\pm 4\pi iu])^{-1}t^\pm}
\hs{0.1 cm}t^\mp_{(n)}\cr
&&={_1\over^2}\sum_{m}(1-\exp[\pm4\pi iu])^{-1}t^\pm_{(m)}
\hs{0.1 cm}\omega(z_n,w|\eta_u)\hs{0.1 cm}
t^{\mp}_{(n)}\mp(1-\exp[\pm 4\pi iu])^{-1}\omega(z_n,
w|\eta_u)\hs{0.1 cm}t^3_{(n)}\hs{0.1 cm}.\non
\qqq

\nappendix{C}
\vskip 0.5cm

\no We will list here,
without proofs, some elementary
facts about the behavior of functions $\omega$ and
$\Pi(u,\tau)$ under products
and modular transformations. They are useful
in studying the connection $\tilde\nabla(w)$.
First let introduce some notations:
\qq
\omega^\pm(z,w|u)&=&
\tr\left[t^\mp\left( \omega(z,w|\eta_u)+{i\over1-\ee^{\pm4\pi
iu}}\right)t^\pm
\right]\s,
\cr
\tilde\omega^\pm(z,w|u)&=&
\tr\left[t^\mp\left( \tilde\omega(z,w|\eta_u)+
{i\ee^{\pm4\pi iu}\over(1-\ee^{\pm4\pi iu})^2}\right)t^\pm
\right]\s,
\cr\cr\rlap{$\tilde\omega(z,w)$}
\phantom{\tilde\omega^\pm(z,w|u)}
&=&
\tilde\omega(z,w|\eta_u=1)\ .
\qqq
One has
\qq
\hspace{-0.7cm}\omega^-
(z_n,{w}|u)\omega^+(z_m,{w}|u)&=&\omega(z_n,{w})\omega^+(z_m,z_n|u)
\ceqno\cr&+&\omega(z_m,{w})\omega^-(z_n,z_m|u)
-i\tilde\omega^+(z_m,z_n|u)
&for $z_m\not=z_n$\s,\hskip .5cm\cr
\qqq
\qq
\hspace{-0.7cm}\omega(z_n,{w})\omega(z_m,{w})&=&
\omega(z_n,{w})
\omega(z_m,z_n)
+
\omega(z_m,{w})\omega(z_n,z_m)
\ceqno\cr\vbox to .5cm{}&-&i
\tilde\omega(z_m,z_n)-i
\tilde\omega(z_m,{w})
-i\tilde\omega(z_n,{w})
& for $z_m\not=z_n$\s.\hskip .5cm\cr
\qqq
The above formulae are (approximate) versions of the identity
$$\bar\partial^{-1}(f\bar\partial^{-1}g)+\bar\partial^{-1}
((\bar\partial^{-1}f)g)=
(\bar\partial^{-1}f)(\bar\partial^{-1}g)
$$
which does not hold exactly because boundary conditions
of $\omega$ are not preserved by the products. On the other hand, one
has
\qq
\omega^+(z_n,{w}|u)\omega^-(z_n,{w}|u)&=&
\partial_{w}\omega(z_n,{w})
-i
\tilde\omega^+(z_n,z_n|u)\s,
\qqq
\qq
\omega(z_n,{w})\omega(z_n,{w})&
=&\partial_{w}\omega(z_n,{w})
-i\omega(z_n,{w})-
i\tilde\omega(z_n,z_n)+2
\tilde\omega(z_n,{w})\s.
\qqq
\vskip 0.2cm

Under shifts and reflections in $u$, quantity $\Pi(u,\tau)$ of
eq. (\ref{bigPi}) transforms by
\qq
\hbox to 2.4cm{$\displaystyle \Pi(u+1/2,\tau)$\hfil}&=
&-\Pi(u,\tau)\s,\cr
\hbox to 2.4cm{$\displaystyle \Pi(u+\tau/2,\tau)$\hfil}&=
&-\ee^{-\pi i (4u+\tau)}\Pi(u,\tau)\s,\cr
\hbox to 2.4cm{$\displaystyle \Pi(-u,\tau)$\hfil}&=
&-\Pi(u,\tau)
\qqq
\vskip -0.2cm
\no and
\vskip -0.2cm
\qq
\hbox to 3.1cm{$\displaystyle\omega^\pm(z,w|-u)$\hfil}
&=& -\omega^\pm(w,z|u)=\omega^\mp(z,w|u)\s,\cr
\hbox to 3.1cm{$\displaystyle\omega^\pm(z,w|{u+1/2})$\hfil}
&=&\omega^\pm(z,w|{u})\s,\cr
\hbox to 3.1cm{$\displaystyle\omega^\pm(z,w|{u+\tau/2})$\hfil}&=&
\ee^{i(z-w)}\omega^\pm(z,w|u)\s.
\qqq
\vskip 0.2cm

Under the transformations by
$\s\left(\pmatrix{a&b\cr c&d},(p_n,q_n)\right)\in \Gamma_N\s$
of (\ref{mod}),
$$\tau\rightarrow(a\tau+b)/(c\tau+d),\ \ \ \
z_n\rightarrow (z_n+2\pi p_n+2\pi\tau q_n)/(c\tau+d),\ \ \ \
u\rightarrow u/(c\tau+d)\s,$$
$\omega$ and $\Pi$ transform as follows
\qq
\omega^\pm(z_n,z_m|u)&\rightarrow&
(c\tau+d)\ \omega^\pm(z_n,z_m|u)
\ee^{\pm2 iu [2\pi (q_n-q_m)-c(z_n-z_m)/(c\tau+d)]}\s\cr
\omega^3(z_n,z_m)&\rightarrow&(c\tau+d)
\left(\omega^3(z_n,z_m)-i(q_n-q_m)+i{{
c(z_n-z_m)\over 2\pi(c\tau+d)}}+{i/2}
\right)-{i/2}\s,
\cr
\Pi(u,\tau)&\rightarrow& \pm(c\tau+d)^{1/2} \ee^{4\pi i c
u^2/(c\tau+d)}
\Pi(u,\tau)\s.
\qqq
The last expression is not uniquely defined because
of the presence of a square
root. This is, however, all we need since $\Pi(u,\tau)$
appears in $\tilde\nabla(w)$ together with its inverse.

\nappendix{D}
\vskip 0.5cm

\no In this Appendix we show how operators $D^L_{{\un j},\alpha}$
commute with connection $\tilde\nabla(w)$ on bundle $W_{\un j}$.
This computation
allows to complete the proof of invariance
of $W_{\un j}^{fr}(\un z,\tau)$ under parallel transport
by the connection.

First we need some simple properties of this operators.
Note that generators of the Lie algebra in the polynomial
realization are
\qq
t^3_{(n)}&=&-v_n\partial_{v_n}+j_n\s,\cr
t^+_{(n)}&=&-\partial_{v_n}\s,\cr
t^-_{(n)}&=&v^2_n\partial_{v_n}+2j_nv_n\s.
\qqq
Then from the fact that $\sum_n t_{(n)}^3 \gamma=0$ for
$\gamma\in W_{\un j}(\un z,\tau)$ one deduces that
\qq
\sum_{n=1}^N D_{{\un j},\alpha}^{L^n}=\sum_{n=1}^N (j_n-l_n) D_{{\un
j},\alpha}^L
\label{prop1}
\qqq
(${L^n}$, resp. ${L_n}$, stands for
the $N$-tuple which has all elements equal to those of
$L$ except for the $n^{\rm th}$ one that is increased,
resp. decreased, by $1$). On the other hand
using $(\ref{cond},d)$ one has, for $|L|-J\in2\NZ+1$,
\qq
2D_{{\un j},\alpha}^L+\sum_n l_n(l_n-1-2j_n)D_{{\un
j},\alpha}^{L_n}=B^{L}_{L'}
D_{{\un j},\alpha}^{L'}
\label{prop2}
\qqq
where $B^L_{L'}\not= 0$ only if $|L'|\leq|L|-2$.
\vskip 0.1cm

Now consider operators $R_n$ of (\ref{rreg}). Due to (\ref{prop1})
\qq
\sum_m t^+_{(m)} \gamma(0)&=& \sum_{L;\atop {l_0=0,\atop  |L|=J-1}}
\left(\prod_m {v^{l_m}/l_m!}\right) D_{{\un j},0}^L\s\gamma\cr
\sum_m t^-_{(m)} \gamma(0)&=& (-1)^J\sum_{L;\atop {l_0=0,\atop
|L|=J-1}}
\left(\prod_m {v^{2j_m-l_m}/l_m!}\right) D_{{\un j},0}^L\s\gamma
\qqq
and finally for $R_0$
\qq
([\sum_m t_m^+,\sum_m t_m^-]-4)\s\partial_u\gamma(0)=
\sum_{L,L'\atop l_0=0;\ |L|=J}
\left(\prod_m{v^{l^m}/l_m!}\right) B^{L^0}_{L'} D_{{\un
j},0}^{L'}\s\gamma
\qqq
where $B$ comes from (\ref{prop2})
i.e. only $|L'|<J$ will give non vanishing
terms. This shows that operators $R_n$ are zero for $\gamma\in W_{\un
j}^{fr}$
as was stated before.

For the rest of the connection we have
\qq
D_{{\un j},\alpha}^L \tilde\nabla({w})\gamma&&=\hspace{0.09cm}
(k+2)\Bigl(\sum_m\omega( z_m,{w}) \p_{ z_m}+{\textstyle{1\over2\pi
i}}
\p_\tau\Bigr)D_{{\un j},\alpha}^L\gamma
\cr&&+{1\over 4\pi l_0}\sum_{m,n}\partial_u\tr( t^- \omega(
z_m,{w}|\eta_u)t^+)
(-2D_{{\un j},\alpha}^{L^{nm}}+2(j_n-l_n)D_{{\un
j},\alpha}^{l^m}\cr&&\hskip 6.3cm
-2\delta_{mn}D_{{\un j},\alpha}^{L^m}-2(l_m-j_m)D_{{\un
j},\alpha}^{l^n})
\gamma
\cr&&-{l_0+2\over 4\pi(l_0+1)}\sum_m\omega( z_m,{w})
(2D_{{\un j},\alpha}^{L^{0m}}+2(l_m-j_m)D_{{\un
j},\alpha}^{L^0})\gamma\cr&&
-{1\over 4\pi(l_0+1)}\sum_{m,n}\omega( z_m,{w})\Bigl(
2(l_n-j_n)D_{{\un j},\alpha}^{L^{0m}}-2(l_m-j_m)D_{{\un
j},\alpha}^{L^{0n}}\cr&&
\hskip 4.7cm
-l_n(2j_n-l_n+1)D_{{\un j},\alpha}^{L^{0m}_n}
+l_m(2j_m-l_m+1)D_{{\un j},\alpha}^{L^{0n}_m}\Bigr)\gamma\cr&&
+{l_0+3\over
(4\pi)^2(l_0+1)}D_{{\un j},\alpha}^{L^{00}}\gamma-{1\over
12(4\pi)^2}\sum_{m,n}
(D_{{\un j},\alpha}^{L^{mn}}+2(l_m-j_m+\delta_{mn})D_{{\un
j},\alpha}^{L^n})\gamma\cr&&
+{1\over(4\pi)^2(l_0+1)(l_0+2)}\sum_{n,m}(D_{{\un
j},\alpha}^{L^{00mn}}+
2(l_m-j_m)D_{{\un j},\alpha}^{L^{00n}}+l_m(l_m-1-2j_m)D_{{\un
j},\alpha}^{L^{00n}_m})\gamma\cr&&
+{1\over(4\pi)^2(l_0+1)(l_0+2)}\sum_{m}(2D_{{\un
j},\alpha}^{L^{00m}}+
2(l_m-j_m)D_{{\un j},\alpha}^{L^{00}})\gamma\cr\cr&&
+\sum_{\beta,L'}A_{\alpha L'}^{\beta L} D_{{\un
j},\beta}^{L'}\gamma\label{DL}
\qqq
where
$\alpha,\beta=0,{1\over2},
{\tau\over2},{\tau+1\over2}$
and here and below $A$ represents
different matrices
with non-zero matrix elements only for
$|L'|\leq{\rm max}(|L|,J-1)$.

With the application of eq. (\ref{prop1}),
previous expression reduces to
\qq
D_{{\un j},\alpha}^L \nabla({w})\gamma&=&
(k+2)\Bigl(\sum_m\omega( z_m,{w}) \p_{ z_m}+
{\textstyle{1\over 2\pi i}}\p_\tau
\Bigr)D_{{\un j},\alpha}^L\gamma
\cr&&
+{1\over 4\pi(l_0+1)}\sum_m\omega( z_m,{w})
\Bigl[(J-|L|-2)(2D_{{\un j},\alpha}^{L^{0m}}+2(l_m-j_m)D_{{\un
j},\alpha}^{L^0})\cr&&
\hskip 4.7cm -\sum_n l_n(l_n-1-2j_n)D_{{\un
j},\alpha}^{L^{0m}_n}\Bigr]\gamma\cr&&
+{1\over (4\pi)^2 (l_0+1)(l_0+2)}\Bigl[
(J-|L|-2)(|L|-J-2l_0-3)D_{{\un j},\alpha}^{L^{00}}
\cr&&\hskip 4cm +(J-|L|+l_0+1)
\sum_m l_m(l_m-1-2j_m)D_{{\un j},\alpha}^{L^{00}_m}\Bigr]\gamma
\cr\cr&&
+A_{\alpha L'}^{\beta L} D_{{\un j},
\beta}^{L'}\s\gamma\s.\label{DLR}
\qqq
Note that for $|L|\leq J-2$ eq. (\ref{DLR}) is equivalent
to (\ref{DLC}).
We must still proof that the latter also holds
for $|L|=J-1$. In this case, or more generally
for $|L|-J+1\in 2\NZ$, property (\ref{prop2}) applies and
one has
\qq
D_{{\un j},\alpha}^L \nabla(w)\gamma&=&
(k+2)\Bigl(\sum_m\omega( z_m,w) \p_{ z_m}+
{\textstyle{1\over 2\pi i}}\p_\tau
\Bigr)D_{{\un j},\alpha}^L\gamma
\cr&&\hskip -2 cm
+{|L|-J+1\over 4\pi (l_0+1)}
\Bigl(\sum_m\omega( z_m,w)\sum_n l_n(l_n-1-2j_n) D_{{\un
j},\alpha}^{L^{0m}_n}
\clabel{odd}\cr
&&\ \ \ \ \ \ \ \ -{J-|L|+2l_0+4\over 4\pi (l_0+2)}
\sum_n l_n(l_n-1-2j_n) D_{{\un j},\alpha}^{L^{00}_n}\Bigr)\gamma
\cr
&&
\hskip -2cm
+A_{\alpha L'}^{\beta L} D_{{\un j},\beta}^{L'}\gamma\s.
\qqq
Writing (\ref{DLR}) and (\ref{odd})
in components, one obtains eqs. (\ref{DLC}).

\end{document}